\documentclass[acmsmall]{acmart}

\acmJournal{TOSEM}

\usepackage[utf8]{inputenc}
\usepackage[T1]{fontenc}

\usepackage{newtxmath} 
\usepackage{array}
\usepackage{multirow}
\usepackage{graphicx}
\usepackage{subfig}
\usepackage{balance}
\usepackage{algorithm}
\usepackage{algorithmicx}
\usepackage{esvect}
\usepackage{booktabs}
\usepackage{tabularx}
\usepackage{anyfontsize}
\usepackage[all]{nowidow}
\usepackage[normalem]{ulem}
\usepackage{xcolor}
\usepackage{colortbl}
\usepackage{enumitem}
\usepackage{makecell}
\usepackage[switch]{lineno}
\usepackage{framed}
\usepackage{wrapfig}
\usepackage{varwidth}
\usepackage{lineno}

\newcolumntype{Y}{>{\centering\arraybackslash}X}

\definecolor{commentGray}{RGB}{120,120,120}
\renewcommand{\algorithmiccomment}[1]{\bgroup\color{commentGray}{//#1}\egroup}

\usepackage{pifont}
\usepackage{listings}
\usepackage{framed}
\usepackage{tikz}
\definecolor{light-gray}{gray}{0.9}
\usepackage[framemethod=TikZ]{mdframed}
\mdfsetup{skipabove=5pt,skipbelow=3pt}
\usepackage{lipsum}
\mdfdefinestyle{RQFrame}{%
	linecolor=black,
	outerlinewidth=0.15pt,
	roundcorner=3pt,
	innertopmargin=2pt,
	innerbottommargin=2pt,
	innerrightmargin=4pt,
	innerleftmargin=4pt,
	backgroundcolor=light-gray}

\newcommand{\setalglineno}[1]{%
  \setcounter{ALG@line}{\numexpr#1}}

\newcommand{\element}[1]{{\text{\fontsize{10}{10}\textsf{#1}}}}

\definecolor{javagreen}{rgb}{0.25,0.5,0.35} 

\lstdefinestyle{Alg}{
  basicstyle=\ttfamily\footnotesize,
  breaklines=true,
  tabsize=2,
  mathescape,
  numbers=left,
  xleftmargin=2.5em,
  xrightmargin=0.5em,
  frame=tb,
  framexleftmargin=2em,
  emph={Algorithm,Input,Output,for,each,do,if,else,Function,while,let,be,repeat,until,return,times,and,or,break,in,then,},
  emphstyle={\textbf},
  escapechar=?,
  morecomment=[l][\color{javagreen}]{//},
  columns=flexible,
}

\newcounter{commentnumber}

\begin{document}

%
\title{Test Generation Strategies for Building Failure Models and Explaining Spurious Failures}
%
%
%
\author{Baharin A. Jodat}
\email{balia034@uottawa.ca}
\affiliation{%
  \institution{University of Ottawa, Canada}
}

\author{Abhishek Chandar}
\email{achan260@uottawa.ca}
\affiliation{%
  \institution{University of Ottawa, Canada}
}

\author{Shiva Nejati}
\email{snejati@uottawa.ca}
\affiliation{%
  \institution{University of Ottawa}, Canada
}

\author{Mehrdad Sabetzadeh}
\email{m.sabetzadeh@uottawa.ca}
\affiliation{%
  \institution{University of Ottawa, Canada}
}

\renewcommand{\shortauthors}{B. A. Jodat, A. Chandar, S. Nejati, M. Sabetzadeh}

\begin{abstract}
Test inputs fail not only when the system under test is faulty but also when the inputs are invalid or unrealistic. 
Failures resulting from invalid or unrealistic test inputs are spurious. Avoiding spurious failures improves the effectiveness of testing  in exercising the main functions of a system, particularly for compute-intensive (CI) systems where a single test execution takes significant time. In this paper, we propose to build failure models for inferring  interpretable rules on test inputs that cause spurious failures. We examine two alternative strategies for building failure models: (1)~machine learning (ML)-guided test generation and (2)~surrogate-assisted test generation.
\emph{ML-guided test generation} infers boundary regions that separate passing and failing test inputs and samples test inputs from those regions. \emph{Surrogate-assisted test generation} relies on surrogate models to predict labels for test inputs instead of exercising  all the inputs. We propose a novel surrogate-assisted algorithm that uses multiple surrogate models simultaneously, and dynamically selects the prediction from the most accurate model.  We empirically evaluate the accuracy of failure models inferred based on surrogate-assisted and ML-guided test generation algorithms. 
Using case studies from the domains of cyber-physical systems and networks, we show that our proposed surrogate-assisted approach generates failure models with an average accuracy of $83$\%, significantly outperforming  ML-guided test generation and two baselines. Further, our approach learns failure-inducing rules that identify genuine spurious failures \hbox{as validated against domain knowledge.} 
\end{abstract}



\maketitle

\section{Introduction}
\label{sec:intro}
Traditionally, software testing has been concerned with finding inputs that reveal failures in the system under test (SUT). However, failures do not always indicate faults in the SUT. Instead, failures may arise because the test inputs are \emph{invalid} or \emph{unrealistic}. For example, in an airplane autopilot system, a failure indicating that the ascent  requirement fails for a test input that points the plane nose downward would be invalid. 
This is because the failure is caused by an unmet environment assumption:
the nose should be upward for  ascent.  For another example, consider a network-management system.  In such a system, quality-of-service requirements will inevitably fail for unrealistic test inputs that  overwhelm the network beyond its capacity.  Environment assumptions capture the expected conditions for a system's operational environment~\cite{giannakopoulou2002assumption}. Attempting to test a system for a more general environment than its expected operational environment may lead to overly pessimistic testing and verification results.
We refer to failures arising from test inputs violating the system's environment assumptions as \emph{spurious failures}. It is often the case that environment assumptions are not fully known for software systems~\cite{10.1145/3270112.3270115}; therefore, it is difficult to determine whether a failure is indeed spurious.

Automated random testing (fuzzing)~\cite{miller1990empirical} becomes more effective in exercising the main functions of a system if the fuzzer  avoids spurious failures~\cite{ gopinath2020abstracting}. Spurious failures particularly pose a challenge for compute-intensive (CI) systems, where a single test execution takes significant time to complete. For CI systems, we want to use the limited testing time budget to generate valid inputs that exercise the system's main functions. A promising approach for identifying spurious failures is to build \emph{failure models}~\cite{ kifetew2017generating, kampmann2020does,kapugama2022human}. Failure models provide conditions that  explain the circumstances of failures and describe  when a failure occurs and when it does not~\cite{kampmann2020does}. Failure models can infer  rules leading to and only to failures. These rules are candidates to be validated against domain knowledge to determine whether the failures that the rules identify are spurious. 


Recent research on synthesizing input grammars~\cite{kifetew2017generating, kulkarni2021learning,bastani2017synthesizing, aschermann2019nautilus, wang2019superion} and abstracting failure-inducing inputs~\cite{kapugama2022human, gopinath2020abstracting, kampmann2020does} aims to understand the circumstances of different failures. These approaches start from an example failure and iteratively generate more tests to learn the input conditions that lead to that failure. The tests are generated via fuzz testing with or without an input grammar. These approaches are geared towards systems with string inputs, where oracles are typically binary (pass/fail) verdicts. However, these approaches are not optimized for systems with numeric inputs, where the inputs are not governed by grammars and where quantitative fitness functions, developed based on system requirements, are used to determine the degree to which test inputs pass or fail. These quantitative fitness functions enable exploration of input space using multiple test-generation heuristics and learning algorithms, resulting in test sets with sufficient information to infer candidate rules for identifying spurious failures.

This paper proposes a framework to infer failure models for compute-intensive (CI) systems with numeric inputs. Examples of such systems include cyber-physical systems (CPS) and network systems.  We follow a data-driven approach and infer failure models by harvesting information from a set of test inputs. To generate such sets, one can use either \emph{explorative} or \emph{exploitative} search methods~\cite{metaheuristicsbook}. The former attempts to sample the entire search space, whereas the latter attempts to sample  the most informative regions of the search space. The challenge with the explorative approach is that we need to collect and execute many test inputs from the search space to determine if they pass or fail. For CI systems, this takes significant time and may become infeasible. The challenge with the exploitative approach is that one needs effective guidance for sampling within large and multi-dimensional search spaces. 


Machine learning (ML) has been used for improving the effectiveness and efficiency of both explorative and exploitative search~\cite{miningassumption, WCET, ben2016testing, matinnejad2014mil, menghi2020approximation, tong2021surrogate, jin2002fitness}.  For explorative search, \emph{surrogate-assisted test generation} relies on ML to predict verdicts for test inputs instead of executing them all~\cite{jin2005comprehensive, ben2016testing, haq2022efficient, matinnejad2014mil}. Using a quantitative surrogate, one can forego system executions when the predicted verdicts remain valid  after offsetting  prediction errors. Otherwise, we execute the SUT and use the results from the executed test inputs to refine the surrogate. As for  exploitative search, \emph{ML-guided  test generation} aims to infer boundary regions that
separate passing and failing test inputs and to subsequently sample test inputs from those regions~\cite{miningassumption,WCET}. 
The intuition is that tests sampled in the boundary regions are more informative for identifying failures and can  be used to further refine the boundary regions. Both approaches provide a set of labelled test inputs from which one can infer failure models using techniques such as  decision-rule learning~\cite{dataminingbook}. Human experts must nonetheless review and validate the resulting rules to determine whether they represent genuine spuriousness. The use of interpretable ML techniques such as decision-rule learning allows failure models to be expressed as easily understandable rules linked to system inputs, making them ideal for human interpretation.

While surrogate-assisted and ML-guided test generation algorithms have been previously used to generate individual test inputs~\cite{ben2016testing, haq2022efficient} , their efficacy in generating failure models remains unexplored. 
Specifically, earlier work strands~\cite{ben2016testing, haq2022efficient} employ machine learning to more effectively steer test generation towards areas within the search space that are likely to contain the most severe failures. In this paper, we use machine learning to devise new test generation algorithms, with the aim of inferring failure models for systems that have numeric inputs.
We evaluate the resulting failure models against those produced by baselines. Our evaluation  answers two main questions: (1)~How accurate are the failure models generated by the surrogate-assisted and ML-guided techniques in predicting failures? (2)~How useful are failure models for identifying spurious failures? We use two kinds of study subjects in our evaluation: (i)~A benchmark of four CPS Simulink subjects with 12 requirements that are non-compute intensive (non-CI).  (ii)~Two industrial CI systems,  one from the CPS and the other from the network domain. In summary, we make the following contributions: 

{\textbf{(1)} We propose a data-driven framework for inferring failure models for systems with numeric inputs including CPS and network systems (Section~\ref{sec:approach}).

\textbf{(2)} We propose a dynamic surrogate-assisted algorithm that uses multiple surrogate models simultaneously during  search, and dynamically selects the prediction from the most accurate model (Section~\ref{subsec:mainloop}). Our evaluation performed based on seven surrogate-model types in the literature~\cite{haq2022efficient, tong2021surrogate, diaz2016review, dushatskiy2021novel} shows that, compared to using surrogate models individually, our dynamic surrogate-assisted algorithm provides the best trade-off between  accuracy and  efficiency by generating datasets that are at least $33\%$ larger while being at least $28\%$ more accurate (RQ1 in Section~\ref{subsec:rq1}).

\textbf{(3)} We compare the accuracy of failure models obtained using our dynamic  surrogate-assisted approach against two ML-guided techniques as well as two baselines. One baseline is random-search, and the other is an adaptation of a state-of-the-art approach that generates failure models for systems with structured inputs~\cite{kampmann2020does}. Our results show that our dynamic surrogate-assisted algorithm yields failure models with an average accuracy, precision, and recall of $83$\%, 72\%, and 88\%, respectively,  significantly outperforming the  ML-guided algorithms and the baselines (RQ2 in Section~\ref{subsec:rq2} and RQ3 in Section~\ref{subsec:rq3}). 

\textbf{(4)}  We demonstrate that failure models built using our dynamic surrogate-assisted algorithm generate useful rules for identifying spurious failures in our CI subjects, as validated by domain knowledge (RQ4 in Section~\ref{subsec:rq4}).

\textbf{(5)} We present  lessons learned based on our findings:  The first lesson summarizes the advantages of using  decision rules for building failure models. The second lesson highlights  the limitations of focusing testing  on finding individual failures and why failure models provide better insights about the effectiveness of testing algorithms.

It is essential for systems to handle all potential inputs including those that violate environment assumptions, and hence, are invalid. Spurious failures caused by invalid test inputs indicate a need for additional safeguards against invalid inputs that may be generated, among other sources, by human operator errors or malfunctioning hardware components, such as inaccurate sensor data.  However, these invalid test inputs do not exercise the core functionality of a system. While ensuring that a given system is safeguarded against invalid inputs is crucial, the inability to identify spurious failures can distort our understanding of the system's capabilities.  This may also lead to misplaced confidence in a testing strategy that reveals numerous failures, yet offers little insight into the system’s primary functions.\

\textbf{Organization.} Section~\ref{sec:motivation} motivates the need for identifying spurious failures. Section~\ref{sec:approach} presents  our data-driven framework for inferring failure models and presents alternative surrogate-assisted and ML-guide algorithms for building failure models. Section~\ref{sec:evaluation} presents an evaluation of these algorithms. Section~\ref{sec:discuss} outlines the main lessons learned from the research. Section~\ref{sec:related} compares with  related work. Section~\ref{sec:conclusion} summarizes the paper and suggests directions for future work.

\section{Motivation}
\label{sec:motivation}
Using two real-world, compute-intensive (CI) systems, we motivate the need for identifying spurious failures. These systems, both of which are open-source, are a Network Traffic Shaping System (NTSS)~\cite{enrich, cakepaper} and an autopilot system~\cite{autopilotbenchmark}. 

NTSS is typically deployed on routers to ensure high network performance for real-time streaming applications such as teleconferencing (e.g., Zoom). Without  NTSS, voice and video packets may be transmitted out of sequence or with delays. As a result, users may experience choppy or freezing voice/video. Due to the increasing remote-working practices, multiple streaming applications may be running at the same time in homes and small-office settings. This has made systematic testing of NTSS essential as a way to ensure that networks meet their quality-of-experience requirements.

NTSS works by dividing the total network bandwidth into classes with different priorities. The higher-priority classes are typically used for transmitting time-sensitive, streaming voice and video.  To test the performance of an NTSS, we assign data flows  with different bandwidth values to different NTSS classes. The purpose is to ensure that NTSS is configured optimally and can maintain good performance even when a high volume of traffic flows through its different classes. When we stress-test an NTSS, no matter how well-designed the NTSS is, we expect the quality of experience to deteriorate and become unacceptable eventually. Test inputs that stress NTSS beyond a certain limit deterministically fail and do not help reveal flaws or suboptimality in the NTSS design. Our approach in this paper infers the limit on the traffic that can flow through different NTSS classes without compromising the quality of experience. For an NTSS setup with eight classes from \element{class0} to \element{class7}, we learn the following rule specifying failing test inputs:

\vspace*{.15cm}
\fbox{\textsf{r1: IF (class5+ class6+ class7 > 0.75 $\cdot$ threshold) THEN FAIL}}
\vspace*{.15cm}

In the above rule, \element{threshold} is the sum of the maximum bandwidths of classes 5, 6, and 7. As we discuss in Section~\ref{subsec:rq4}, we validate Rule \element{r1}  with a domain expert and confirm that failures specified by this rule are indeed spurious.
Rule \element{r1} indicates that attempting to simultaneously utilize classes 5, 6 and 7 more than  $75$\% of their maximum ranges would compromise  quality of experience for the entire network.  Rule \element{r1}  helps domain experts in at least two ways: (1)~it informs them that test inputs that satisfy the rule are spurious, since such test cases do not reveal  design faults, and (2)~it provides experts with data-driven evidence that the cumulative utilization of classes 5, 6, and 7 should be kept below the identified limit.

Our second case study is an autopilot system from  Lockheed Martin's benchmark of challenge Simulink models~\cite{chaturvedi2017modeling}. This autopilot  system is expected to satisfy the following requirement:\linebreak $\varphi$= ``\emph{When the autopilot is enabled, the aircraft  should reach the desired altitude within 500 seconds in calm air}''. When we test the autopilot by fuzzing,  we find several test inputs that violate this requirement and several test inputs that satisfy it. It is however unclear whether the failures are due to faults in the system or due to missing or unknown assumptions on the system inputs. Failures caused by  missing or unknown assumptions  would be  spurious. 
As we discuss in Section~\ref{subsec:rq4}, we identify the following rule as one that indicates spurious failures: 

\vspace*{.15cm}
\fbox{\fontsize{10}{10}\textsf{r2: IF (PitchWheel[0..300] $\leq$ -28 $\wedge$  Throttle[0..300]  $\leq$ 0.1) THEN FAIL}}
\vspace*{.15cm}

Here, \element{Throttle[0..300]} is the boost applied to the engine by the pilot during the first $300$s, and \element{PitchWheel[0..300]} is the upward or downward degree of  the aircraft nose, again during the first $300$s. Note that both \element{Throttle} and \element{PitchWheel} are signals over time.  In order to validate \element{r2}, we  examined the  handbook of the De Havilland Beaver aircraft~\cite{autopilothandbook}.  According to the handbook,  for this aircraft type, to satisfy requirement $\varphi$, the pilot should manually adjust the throttle boost (\element{Throttle}) to a sufficiently high value. The handbook further states that to be able to ascend, the plane's nose should not be pointing downward. That is, \element{r2} describes a situation where the pre-conditions for $\varphi$ are not met. Hence, the tests in these ranges are expected to fail and  are uninteresting for revealing system faults. Further, \element{r2}  can be used for implementing safeguards against misuse by the human operator (pilot).

\section{Generating Failure Models}
\label{sec:approach}
Figure~\ref{fig:fig1} shows our framework for generating failure models.  The inputs to our framework are: (1)~an executable system or simulator $S$, (2)~the input-space representation $\mathcal{R}$ for $S$, and (3)~a quantitative fitness function $F$ for each requirement of $S$; an example requirement, $\varphi$, for autopilot was given in Section~\ref{sec:motivation}. The full set of requirements for our case studies is available in our supplementary material~\cite{requirements}. We make the following assumptions about the search input space ($\mathcal{R}$) and the fitness function ($F$):

\begin{itemize}
\item \textbf{A1}~We assume that the system inputs  are variables of type real or enumerate. For each real variable, the range of the values that the variable can take is bounded by an upper bound and a lower bound.  

\item \textbf{A2}~For each requirement of $S$, we have a fitness function $F$ based on  which a pass/fail verdict can be derived for any test input. Further, the value of $F$ differentiates among the pass test inputs, those that are more acceptable, and among the fail test inputs, those that trigger more severe failures.  Specifically, we assume that the range of $F$ is an interval $[-a, b]$ of $\mathbb{R}$. For a given test,  $F(t) \geq 0$  iff $t$ is passing, and otherwise, $t$ is  failing. The closer $F(t)$ is to $b$, the higher the confidence that $t$ passes; and the closer  $F(t)$ is to $-a$, the higher the confidence that $t$ fails.  
\end{itemize}
Assumptions \textbf{A1} and \textbf{A2} are common for CPS models expressed in Simulink~\cite{chaturvedi2017modeling,menghi2019generating, tuncali2019requirements,arrieta2019pareto},  automated driving systems~\cite{haq2022efficient}, and network-management systems~\cite{enrich}, and are valid for all the case studies we use in our evaluation.





\begin{figure}[t]
    \centering
    \includegraphics[width=\columnwidth]{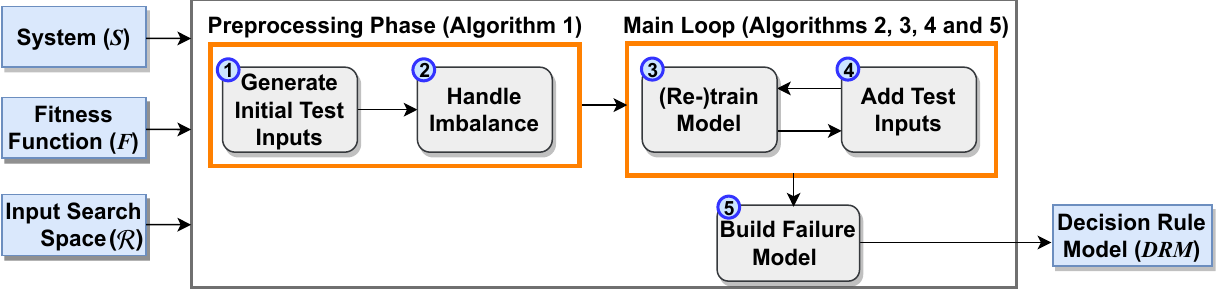}
     
    \caption{Our  framework for generating failure models. 
     The main loop of the framework, i.e., steps 3 and 4, can be realized using  two alternative test-generation strategies: (1)~Surrogate-assisted test generation, or (2)~ML-guided test generation. For  surrogate-assisted test generation, one can use either Algorithm~2, which is based on an individual surrogate model, or Algorithm~3, which is based on our proposed dynamic model. For  ML-guided test generation, one can use either Algorithm 4, which uses 
    regression trees to identify the boundary regions between passing and failing test cases, or Algorithm 5, which uses logistic regression for the same purpose.}\label{fig:fig1}
     
\end{figure}



As discussed in Section~\ref{sec:intro}, we examine two alternative test-generation approaches for building failure models: \emph{surrogate-assisted} and \emph{ML-guided}. Both approaches can be captured using  the framework shown in Figure~\ref{fig:fig1}:  The preprocessing phase generates a set of test inputs labelled with fitness values. The main loop takes the test-input set created by the preprocessing phase, and trains a model. When the framework is instantiated for surrogate-assisted test generation, the model predicts fitness values for the generated test inputs. When the framework is instantiated for ML-guided test generation, the model guides  test-input sampling.  The main loop extends the test-input set using the trained model while also refining the model based on  newly generated tests. After the main loop terminates, the framework uses the test-input set to train, using decision-rule learners, a failure model. In the remainder of this section, we detail each step of  the framework shown in Figure~\ref{fig:fig1}. 



\subsection{Preprocessing Phase}
\label{subsec:pre}




\begin{algorithm}[t]

\caption{The preprocessing phase of Figure~\ref{fig:fig1}}
\label{alg:alg1}
\begin{flushleft}

\textbf{Input} $S$: System\\
\textbf{Input} $\mathcal{R} = \{R_1, \ldots, R_n\}$: Ranges $\mbox{for}$ input variables $v_1$ to $v_n$\\
\textbf{Input} $F$: Fitness $\mbox{Function}$\\
\textbf{Input} $d$: The budgeted dataset size\\
\textbf{Output} $\mathit{DS}$: A set of test inputs $\mbox{and}$ their fitness values \\
\end{flushleft}
\begin{algorithmic}[1]
    
\State \hspace{-0.1cm} $\mathit{DS}^i$ $\leftarrow$  GenerateTests($\mathcal{R}$, $d/2$); \Comment{(Adaptive) Random Testing}
\State \hspace{-0.1cm} $\mathit{DS}^l \leftarrow$ Execute($S$, $\mathit{DS}^i$, $F$); \Comment{Compute fitness values}
\State \hspace{-0.1cm} $\mathit{DS}^b$ $\leftarrow$ HandleImbalance($\mathit{DS}^l$); \Comment{Use SMOTE to generate synthetic test inputs}
\State \hspace{-0.1cm} $\mathit{DS}^{b,l}$ $\leftarrow$ Execute($S$, $\mathit{DS}^b$, $F$); \Comment{Compute fitness values of the tests generated by SMOTE}
\State \hspace{-0.1cm} $DS$ $\leftarrow$ $\mathit{DS}^{b,l} \cup \mathit{DS}^l$; \Comment{Combine the test inputs generated by adaptive random testing and SMOTE along with their fitness values  to form a dataset}
\State \hspace{-0.1cm} \textbf{return} $\mathit{DS}$

\end{algorithmic}

\end{algorithm}
 Algorithm 1 describes the preprocessing phase that generates the initial dataset  for training a model to be used in the main loop of Figure~\ref{fig:fig1}. This algorithm first randomly generates half ($d/2$) of the budgeted test inputs and computes a fitness value for each test input by executing the test input using $S$ (lines~1-2). Since ML models perform poorly when the training set is imbalanced, we attempt to address any potential imbalance  before using the data for ML training~\cite{dataminingbook} (line~3).  In our work, the imbalance, if one exists, is between the pass and the fail classes. We use the well-known synthetic minority over-sampling technique (SMOTE) for addressing  imbalance~\cite{smote}. 
Let $\mathit{minor}$ (resp. $\mathit{major}$) be  the number of tests in the minority (resp. majority) class. SMOTE over-samples the minority class 
by taking each minority-class sample and introducing synthetic examples along the line segments joining any/all of the $k$ minority-class nearest neighbours~\cite{smote}. The process is repeated until we have $m= \mathit{major} - \mathit{minor}$ new such tests.

We discard the labels from SMOTE  and instead execute the tests to compute their actual  fitness values (line~4). 
We discard the labels provided by SMOTE, since SMOTE categorizes test inputs as pass or fail. Instead, we require test inputs to be labelled with their quantitative fitness values; this enables us to train regression ML models in Step~$3$ of our approach in Figure~\ref{fig:fig1}.
The final dataset ($\mathit{DS}$) is returned at the end (line~6). Although not shown in Algorithm 1, to have exactly $d$ tests in $\mathit{DS}$, we generate the remaining $(d/2 - m)$ tests randomly. Generating these remaining tests randomly does not introduce a new imbalance problem because if random test generation leads to major imbalance, then $m$ is already close to $d/2$ and only a few additional tests need to be generated. Otherwise, a small $m$ indicates that random testing is relatively balanced; in that case, no special provision is necessary for imbalance mitigation in the randomly generated datatset. Since we discard the labels generated by SMOTE and compute the actual labels, the imbalance problem may in principle persist even after applying SMOTE. For our experiments  (Section~\ref{sec:evaluation}), most synthetic samples generated by SMOTE indeed belong to the minority class. Our preprocessing therefore successfully addresses imbalance in our case studies.



\subsection{Main Loop}
\label{subsec:mainloop}
The main data-generation loop is realized via two alternative algorithms, described below: surrogate-assisted and ML-guided.

The goal of this approach is to use surrogates to predict fitness values for some test inputs and thus not execute the system for all test inputs. Hence, surrogates help explore a larger portion of the input space and generate larger test sets.

\subsubsection{Surrogate-Assisted Test Generation.} 
\label{sec:surrogate}
Figure~\ref{fig:figalg2} illustrates the surrogate-assisted test generation process, which takes the same inputs as the framework in Figure~\ref{fig:fig1}. The output is a labelled dataset, $\mathit{DS}$, used in Step~$5$ of Figure~\ref{fig:fig1} to build failure models. The procedure in Figure~\ref{fig:figalg2} is as follows:

\begin{enumerate}
\item  Start with preprocessing (Algorithm~\ref{alg:alg1}) to produce an initial dataset (Step~$1$ of Figure~\ref{fig:figalg2}).
\item  Train a surrogate model with the initial dataset (Step~$2$ of Figure~\ref{fig:figalg2}).
\item  Generate a new test input (Step~$3$ of Figure~\ref{fig:figalg2}) and predict its fitness value using the surrogate model (Step~$4$ of Figure~\ref{fig:figalg2}).
\item  Calculate a confidence interval for the predicted fitness value (Step~$5$ of Figure~\ref{fig:figalg2}).
\item  If the prediction is not accurate, run system $S$ to obtain the actual fitness value (Step~$7$ of Figure~\ref{fig:figalg2}), add the test input along with its actual fitness value to the dataset (Step~$8$ of Figure~\ref{fig:figalg2}), and return to Step~$2$ of Figure~\ref{fig:figalg2} to re-train the surrogate model.
\item  If the prediction is accurate; add the test input along with its predicted fitness value to the dataset (Step~$6$ of Figure~\ref{fig:figalg2}); and, return to Step~$3$ of Figure~\ref{fig:figalg2}.
\end{enumerate}
Note that if we execute system $S$ for a test input, as mentioned in item~$(5)$ above, the surrogate model is retrained using  the dataset updated with this new test execution. In contrast, if system execution is not required, as in item~$(6)$ above, retraining the surrogate model is unnecessary. \\

\begin{figure}[t]
    \includegraphics[width=\columnwidth]{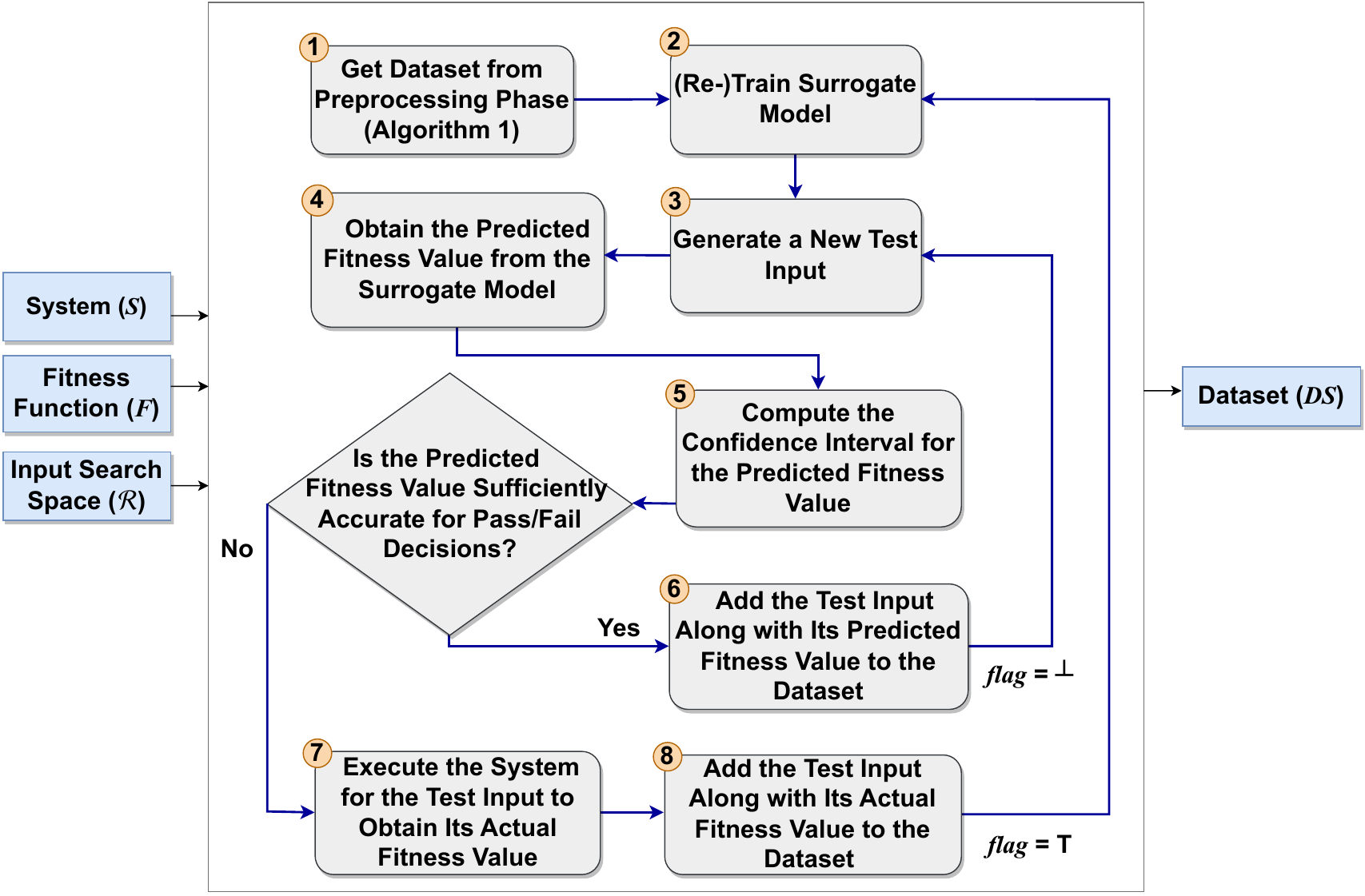}
    \caption{Illustration of the work flow of Algorithm 2 }\label{fig:figalg2}
\end{figure}

To demonstrate the calculation of the confidence interval described in item~$(4)$ above,  consider the example provided in Figure~\ref{fig:fitnessshift}. In this figure, two sample test inputs, denoted as $t$ and $t'$, are shown.
Suppose the predicted fitness values are $\bar{F}(t) = 8$ and $\bar{F}(t') = -1$, indicating a pass label for $t$ and a fail label for $t'$, respectively. Assume that the prediction error, $e$, of the surrogate model is $2$. The confidence interval is calculated as $\bar{F} \pm e$.  That is, the confidence interval for $\bar{F}(t)$ is $[6, 10]$, while the confidence interval for $\bar{F}(t')$ is $[-3, 1]$. 
Using the confidence intervals, we determine whether to execute system $S$ for $t$ and $t'$. For input $t$, the confidence interval of $\bar{F}(t)$ falls entirely within the positive range. Hence, even after accounting for error, we still label the test input $t$ as a pass. Therefore, there is no need for system $S$ to be executed for input $t$, since $t$ can be confidently labelled as pass (item~$(6)$ above). However, for input $t'$ the confidence interval of $\bar{F}(t')$ spans both positive and negative values. This indicates that a label cannot be confidently assigned to $t'$. As a result, system $S$ needs to be executed for $t'$ to obtain its actual fitness value and an accurate verdict (item~$(5)$).
\begin{figure}[t]
    \includegraphics[width=0.9\columnwidth]{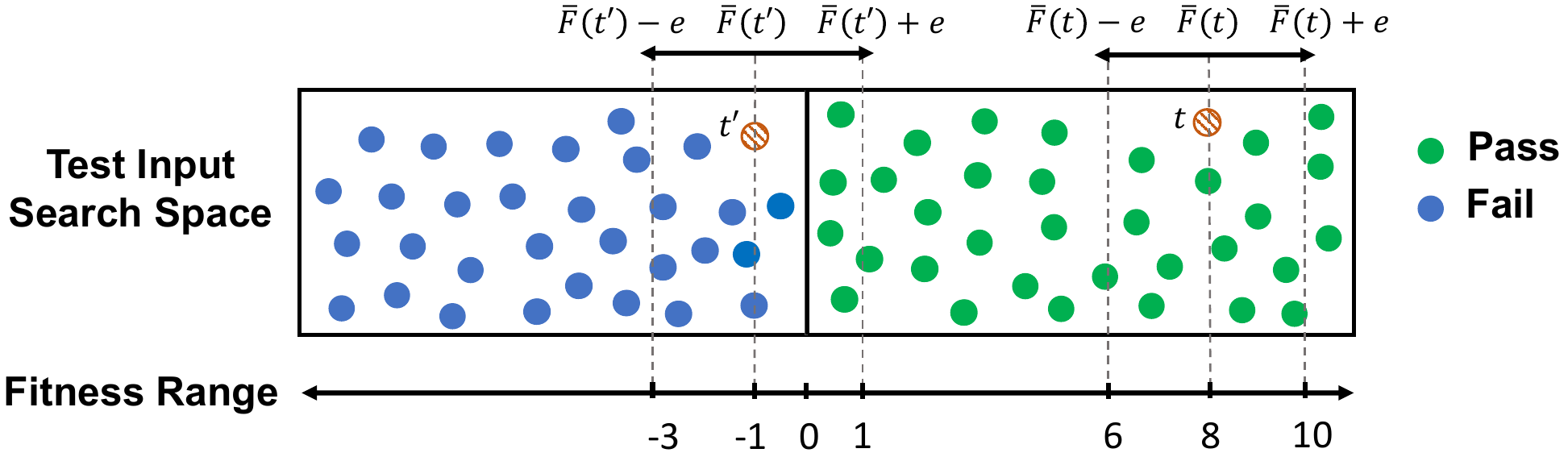}
     
    \caption{Illustration of how to calculate the confidence interval for predicted fitness values to determine whether a predicted fitness value is accurate. Specifically, the figure shows confidence intervals, i.e., $\bar{F} \pm e$, of the predicted fitness values for two  test inputs $t$ and $t'$. For the test input $t$ where the predicted fitness confidence interval remains entirely in the positive range (i.e., the verdict of the test inputs in this range is pass), we do not execute the system since we can say that the test input is passing even after accounting for error $e$. However, for the test input $t'$, the predicted fitness confidence interval spans both positive and negative values (i.e., it includes test inputs with both pass and fail verdicts). Therefore, we need to execute the system for $t'$ to obtain its actual fitness value.}\label{fig:fitnessshift}
     
\end{figure}


The pseudo code of the surrogate-assisted test generation is provided in Algorithm~\ref{alg:alg2}. Below, we discuss each line of this algorithm in detail.

\begin{algorithm}[t]

\caption{Test generation using surrogates}
\label{alg:alg2}
\begin{flushleft}
\textbf{Input} $S$: System \\
\textbf{Input} $\mathcal{R} = \{R_1, \ldots, R_n\}$: Ranges $\mbox{for}$ input variables $v_1$ to $v_n$\\
\textbf{Input} $F$: Fitness $\mbox{Function}$\\
\textbf{Input} $\textsc{InitialDatasetSize}$: size of initial dataset\\
\textbf{Output} $\mathit{DS}$: A dataset to train failure models \\
\end{flushleft}
\begin{algorithmic}[1]
\State \hspace{-0.1cm} $\mathit{DS}\leftarrow $  Preprocessing($S$, $\mathcal{R}$, $F$, $\textsc{InitialDatasetSize}$); \Comment{Run Algorithm~$1$ to generate an initial dataset}
\State $\mathit{DS}^l\leftarrow \mathit{DS}$; $\mathit{flag} \leftarrow $  $\top$; 
\State \hspace{-0.1cm} \textbf{while} 
(execution budget remains) \textbf{do} 
\State \hspace{0.1cm} \textbf{if} ($\mathit{flag}$)
\State \hspace{0.1cm}\hspace{0.1cm} \hspace{-0.1cm} $(\mathit{SM}$, $e$) $\leftarrow$ Train($\mathit{DS}^l$);  \Comment{Training $\mathit{SM}$; $e$ is the error}
\State \hspace{0.1cm} $\mathbf{end}$
\State \hspace{0.1cm} $\mathit{t} \leftarrow$ GenerateTests($\mathcal{R}$, $1$); \Comment{Generate one test input}
\State \hspace{0.1cm} $\bar{F}(t) \leftarrow $ $\mathit{SM}_{} $($\mathit{t}$); \Comment{ Use $\mathit{SM}$ to predict a fitness value for $t$}
\State \hspace{0.1cm} \textbf{if} ($\exists \sim \in \{\geq, <\}\cdot(\bar{F}(t) \sim 0) \wedge (\bar{F}(t) \pm e \sim 0)$ \Comment{Calculate the confidence interval of the predicted fitness value to decide whether system $S$ needs to be executed for $t$ or not}
\State \hspace{0.1cm} \hspace{0.1cm} $\mathit{DS} \leftarrow \mathit{DS} \cup \{\langle \mathit{t}, \bar{F}(t) \rangle \}$; \Comment{Add the test input along with its predicted fitness value to the dataset}  
\State \hspace{0.1cm} \hspace{0.1cm} $\mathit{flag} \leftarrow \bot$; \Comment{No SM re-training is needed}
\State \hspace{0.1cm} $\mathbf{else}$ 
\State \hspace{0.1cm} \hspace{0.1cm} $\{\langle t, F(t) \rangle\} \leftarrow $ Execute($S$,$\{t\}$, $F$); \Comment{Obtain the actual fitness value of $t$ by executing system $S$}
\State \hspace{0.1cm} \hspace{0.1cm} $\mathit{DS}^l \leftarrow \mathit{DS}^l \cup \{\langle \mathit{t}, F(t) \rangle \}$; \Comment{Add the test input along with its actual fitness value to the dataset}
\State \hspace{0.1cm} \hspace{0.1cm} $\mathit{DS} \leftarrow \mathit{DS} \cup \{\langle \mathit{t}, F(t) \rangle \}$; $\mathit{flag} \leftarrow \top$; \Comment{SM re-training is needed}
\State \hspace{0.1cm} $\mathbf{end}$
\State \hspace{-0.1cm} $\mathbf{end}$ 
\State \hspace{-0.1cm} \textbf{return} $\mathit{DS}$

\end{algorithmic}
\end{algorithm}

Line~$1$ of Algorithm~\ref{alg:alg2} uses Algorithm~\ref{alg:alg1} to obtain an initial dataset, $\mathit{DS}$: \\[.5em]
\vspace*{.15cm}
\fbox{1: $\mathit{DS}\leftarrow $  Preprocessing($S$, $\mathcal{R}$, $F$, $\textsc{InitialDatasetSize}$);}
\vspace*{.15cm}\\
On line~$2$, $\mathit{DS}^l$ is created to maintain a record of the test inputs for which $S$ is executed. We use this dataset to train a surrogate model $\mathit{SM}$ and compute its error $e$ (line~$5$):\\[.5em]
\vspace*{.15cm}
\fbox{\begin{varwidth}{\textwidth}
2: $\mathit{DS}^l\leftarrow \mathit{DS}$; $\mathit{flag} \leftarrow $  $\top$;\\
5: $(\mathit{SM}$, $e$) $\leftarrow$ Train($\mathit{DS}^l$);
\end{varwidth} }
\vspace*{.15cm}\\
Specifically, we train $\mathit{SM}$ using $80\%$ of $\mathit{DS}^l$ and compute the mean absolute error of $\mathit{SM}$ on the remaining $20\%$ of $\mathit{DS}^l$. The split ratio is based on the well-known 80/20 rule~\cite{patgiri2019empirical}. We employ a boolean variable, $\mathit{flag}$, to decide whether training a surrogate model is necessary. Initially, on line $2$, we initialize $\mathit{flag}$ to $\top$ to ensure that a surrogate model is trained during the first iteration. Then, on line $7$, we randomly generate a new test input and denote it by $t$:\\[.5em]
\vspace*{.15cm}
\fbox{7: $ \mathit{t} \leftarrow$ GenerateTests($\mathcal{R}$, $1$);}
\vspace*{.15cm}\\
Then, on line~$8$, the surrogate model $\mathit{SM}$ predicts a fitness value for $t$. The predicted fitness value is denoted by $\bar{F}(t)$:\\[.5em]
\vspace*{.15cm}
\fbox{8: $\bar{F}(t) \leftarrow $  $\mathit{SM}_{} $($\mathit{t}$);}
\vspace*{.15cm}\\
On line 9, we calculate a confidence interval for $\bar{F}(t)$ based on the prediction error $e$. Specifically, we compute the interval $[\bar{F}(t) - e, \bar{F}(t) + e]$ as the confidence interval for $\bar{F}(t)$. If the confidence interval remains entirely within the positive or negative range,  indicating a definite pass or fail verdict for the test input $t$, we skip executing system $S$ for the test input $t$. This is because we can confidently label it as either pass or fail. 
Subsequently, we add the test input $t$ along with its predicted fitness value (i.e., $\bar{F}(t)$) to $\mathit{DS}$ (line 10). In addition, we set $\mathit{flag}$ to $\bot$ in order to prevent retraining the surrogate model $\mathit{SM}$ in the next iteration (lines 11):\\[.5em]
\vspace*{.15cm}
\fbox{\begin{varwidth}{\textwidth}
9: \textbf{if}$ \ \exists \sim \in \{\geq, <\}\cdot(\bar{F}(t) \sim 0) \wedge (\bar{F}(t) \pm e \sim 0)$\\
10: \ $\mathit{DS} \leftarrow \mathit{DS} \cup \{\langle \mathit{t}, \bar{F}(t) \rangle \};\\ $11:$ \ \mathit{flag} \leftarrow \bot$;
\end{varwidth} }
\vspace*{.15cm}\\
Otherwise, on line~$12$, if the confidence interval spans both positive and negative numbers, indicating that it covers test inputs with both pass and fail verdicts, we proceed to execute system $S$ to calculate the actual fitness value for $t$. Then, 
we add $t$ along with its actual fitness value to $\mathit{DS}$ and $\mathit{DS}^l$ (lines~$14$-$15$).
In this case, we set $\mathit{flag}$ to $\top$ to indicate that a system execution has taken place (line~$15$):\\[.5em]
\vspace*{.15cm}
\fbox{\begin{varwidth}{\textwidth}
12: \textbf{else} \\
13: \ $\{\langle t, F(t) \rangle\} \leftarrow $ Execute($S$,$\{t\}$, $F$); \\ 14: \ $\mathit{DS}^l \leftarrow \mathit{DS}^l \cup \{\langle \mathit{t}, F(t) \rangle \}$;\\
15: $\ \mathit{DS} \leftarrow \mathit{DS} \cup \{\langle \mathit{t}, F(t) \rangle \}$; $\mathit{flag} \leftarrow \top$;
\end{varwidth} }
\vspace*{.15cm}\\
 In the latter case (i.e., lines~$12$-$15$), we re-train the surrogate model $\mathit{SM}$ using the dataset $\mathit{DS}^l$ which includes the new test input and its actual fitness value (lines~4-5).  The algorithm returns the final dataset $\mathit{DS}$ when the execution budget runs out (line~$18$):\\[.5em]
 \vspace*{.15cm}
\fbox{18: \textbf{return} $\mathit{DS}$;}
\vspace*{.15cm}\\
 The execution budget expires when either the system has been executed to the point where the size of $\mathit{DS}^l$ reaches its desired limit, or our time budget is exhausted, depending on which occurs first.

In our experimentation (Section~\ref{sec:evaluation}), we consider the  surrogate-model types  shown in Table~\ref{tab:surrogatemodels}. These surrogate-model types are the most widely used ones in the evolutionary search and software testing literature~\cite{haq2022efficient, tong2021surrogate, diaz2016review, dushatskiy2021novel}. 
As suggested by the literature and also as we  show in our evaluation (Section~\ref{subsec:rq1}), no surrogate-model type consistently outperforms the others~\cite{xu2021ensemble, friese2016building}.  Therefore, it is recommended to use a combination of surrogate models~\cite{HONG2022114835}. In this paper, we propose, to our knowledge, a novel variation of Algorithm 2 where we train multiple surrogate models and use for predicting  fitness values the model that has the lowest error. This variation is shown in Algorithm~3 where we change line~5 of Algorithm 2 to train and tune a list of surrogate models instead of just one model. We then select the surrogate model with the lowest error for making predictions until the next time we re-train the models.  Similarly, each time we execute line~5 of Algorithm 2,  we re-train a list of surrogate models and select the one that has the lowest error. We refer to our proposed variation as \emph{dynamic} surrogate-assisted test generation. 

In both Algorithm 2 and the variation suggested in Algorithm 3, the first time we train a surrogate model, we also tune its hyperparameters using Bayesian optimization~\cite{bayesian}.  We use the same tuned hyperparameters in all future iterations.  The cost of training and tuning surrogate models for the first time is on the same scale as the cost of a single execution of our CI systems. The time for subsequent re-training of surrogate models  is nonetheless negligible since re-training does not involve any tuning. As we discuss in Section~\ref{sec:evaluation}, the overhead of re-training surrogate models does not deteriorate performance compared to other alternatives.



\begin{table}[t]
\caption{Surrogate models and their descriptions.}
\vspace*{-.3cm}
\label{tab:surrogatemodels}
\scalebox{0.75}{
\begin{tabular}{|l|l||l|l|}
\hline
\textbf{Name} & \textbf{Description}                                                                                                     & \textbf{Name}        & \textbf{Description}                       \\ \hline
GL         & \begin{tabular}[c]{@{}l@{}}Gaussian Process Regression -- \\ nonparametric Bayesian with linear kernel.\end{tabular}     & RT                   & regression tree.                           \\ \hline
GNL        & \begin{tabular}[c]{@{}l@{}}Gaussian Process Regression -- \\ nonparametric Bayesian  with nonlinear kernel.\end{tabular} & RF                   & random forest.                             \\ \hline
LSB           & \begin{tabular}[c]{@{}l@{}}Gradient Boosting  -- an ensemble of regression trees.\end{tabular}                        & \multirow{2}{*}{SVR} & \multirow{2}{*}{Support Vector Regression.} \\ \cline{1-2}
NN            & a two-layer feedforward Neural Network.                                                                                  &                      &                                            \\ \hline
\end{tabular}}
\end{table}

\begin{algorithm}[t]

\caption{Dynamically selecting surrogates}
\label{alg:alg3}
\begin{algorithmic}[1]
\setalglineno{3}
\State \hspace{-0.1cm} $\ldots$
\State \hspace{-0.1cm} \textbf{for} i = 1 to $\mathit{sm}$ \textbf{do} \Comment{Train surrogate models $ \mathit{SM}_1, ..., \mathit{SM}_{\mathit{sm}}$}
\State \hspace{0.1cm} ($\mathit{SM}_i$, $e_i$) $\leftarrow$ Train($\mathit{DS}^l$); 
\State \hspace{-0.1cm} $\mathbf{end}$ 
\State \hspace{-0.1cm} $(\mathit{SM}, e) \leftarrow $ Select $\mathit{SM} \in \{\mathit{SM}_1, \ldots, \mathit{SM}_{\mathit{sm}}\}$ with the lowest error $e$
\State \hspace{-0.1cm}$ \ldots$
\end{algorithmic}

\end{algorithm}

\subsubsection{ML-Guided Test Generation}
\label{subsec:mlguided}
ML-guided test generation uses ML models for identifying the boundary regions that discriminate pass and fail test inputs and iteratively concentrating test-input sampling to those regions. The idea is that, irrespective of the separability of the set of test inputs, ML models can shift the focus of sampling from the homogeneous regions where either fail or pass verdicts are scarce to regions where neither fail nor pass would be dominant.  We consider two alternative ML models that can help us sample from such boundary regions: regression trees (Algorithm 4) and logistic regression (Algorithm 5). As we describe below, a regression tree approximates pass-fail boundaries in terms of predicates over inputs variables, while logistic regression infers a linear formula over input variables. 

\textbf{Regression-Tree Guided Test Generation. }
Algorithm 4 uses the $\mathit{DS}$ dataset obtained from Algorithm 1 (the preprocessing phase) to train a regression-tree model (lines~1-3). In our regression-tree models, tree edges are labelled with predicates  $v_i \sim c$ such that $v_i$ is an input variable, $c \in \mathbb{R}$ is a constant and $ \sim\ \in \{\leq, >\}$. The tree leaves partition the given dataset into subsets such that  information gain is maximized~\cite{dataminingbook}. Each leaf is labelled with the average of the fitness measures of the test inputs in that leaf.  Provided with a regression tree, Algorithm 4 identifies predicates  $\{v_{i_1} \sim c_{i_1}, \ldots, v_{i_m} \sim c_{i_m}\}$
that appear on the two paths whose leaf-node values are closest to zero (one above and one below zero). These predicates specify the boundary between pass and fail, and, as such, we call them \emph{boundary} predicates. By simplifying the boundary predicates, each variable can have at most one upper-bound predicate ($v \leq c$) and at most one lower-bound predicate ($v > c$). For each predicate $v_{i_j} \sim c$ where $\sim \in \{\leq, >\}$, the algorithm replaces the existing range $R_{i_j}$ of $v_{i_j}$ with $R'_{i_j} = [c - $5$\%\cdot c, c + $5$\%\cdot c]$ (lines~5-7). This will ensure that we sample $v_{i_j}$ within  the $5$\% margin around the constant $c$. The variables that do not appear in the boundary predicates retain their range from the previous iteration. We note that if $R'_{i_j}$ does not reduce the range for $v_{i_j}$, i.e., the size of $R'_{i_j}$ is greater than $R_{i_j}$, we do not replace $R_{i_j}$ with $R'_{i_j}$. This is to ensure that larger ranges are not carried over to the next iteration if the range has already been narrowed at some previous iteration.
Next, the algorithm  generates a test input  within the constrained search space (line~8), executes the test input, and adds it along with its fitness measure to $\mathit{DS}$ (line~9). The algorithm returns the final dataset after the execution budget runs out  (line~11). The execution budget expires when either the system has been executed to the point where the size of $DS$ reaches its desired limit, or our time budget is exhausted, depending on which occurs first.




To illustrate range reduction for variables using regression trees, consider the example in Figure~\ref{fig:regtreeexample}.  We choose the two thicker paths highlighted in blue since their leaf-node values are closest to zero. 
Variables $v_1$, $v_2$ and $v_3$ appear on these paths. Suppose the initial ranges for $v_1$, $v_2$ and $v_3$ to be $[0, 20]$, $[10, 30]$, and $[1, 7]$, respectively. Using the regression tree and the process described above, the new reduced ranges for $v_1$, $v_2$ and $v_3$ are $[9.5, 10.5]$, $[19, 21]$ and  $[4.75, 5.25]$ respectively. 
By sampling within these ranges, we get to focus test-input generation on  the pass-fail border identified by the regression tree.
\begin{figure}[t]
\includegraphics[width=0.4\columnwidth]{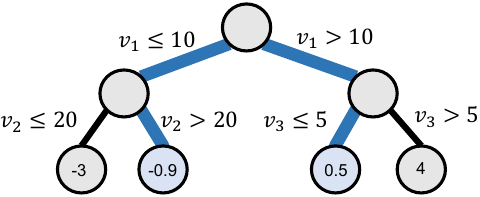}
\caption{A regression tree trained in an iteration of Algorithm~\ref{alg:alg4}. The figure illustrates two regression tree paths chosen for test generation on line~4 of Algorithm~\ref{alg:alg4}.}
\label{fig:regtreeexample}
\end{figure}

\begin{algorithm}[t]
\caption{ML-guided test generation with regression trees}
\label{alg:alg4}
\begin{flushleft}
\textbf{Input} $S$: System\\
\textbf{Input} $\mathcal{R} = \{R_1, \ldots, R_n\}$: Ranges $\mbox{for}$ input variables $v_1$ to $v_n$\\
\textbf{Input} $F$: Fitness $\mbox{Function}$\\
\textbf{Input} $\textsc{InitialDatasetSize}$: size of initial dataset\\
\textbf{Output} $\mathit{DS}$: A dataset to train failure models \\
\end{flushleft}
\begin{algorithmic}[1]
\State \hspace{-0.1cm} $\mathit{DS}\leftarrow $  Preprocessing($S$, $\mathcal{R}$, $F$, $\textsc{InitialDatasetSize}$);\Comment{Run Algorithm~$1$ to generate an initial dataset}
\State \hspace{-0.1cm} \textbf{while} 
(execution budget remains) \textbf{do} 
\State \hspace{0.1cm} $\mathit{RegTree}$ $\leftarrow$ Train($\mathit{DS}$);
\State \hspace{0.1cm} Let $R'_{i_1}$,..., $R'_{i_m}$ be reduced ranges obtained from $\mathit{RegTree}$;
\State \hspace{0.1cm} \textbf{for} each variable $v_{i_j}$ s.t.  $j \in \{1, \ldots, m\}$ \textbf{do}
\State \hspace{0.1cm} \hspace{0.1cm} $\mathcal{R} \leftarrow (\mathcal{R} \setminus \{R_{i_j}\}) \cup \{R'_{i_j}\}$; \Comment{ Replace the range $R_{i_j}$ of $v_{i_j}$ in   $\mathcal{R}$ with the new reduced range $R'_{i_j}$ from line 4}
\State \hspace{0.1cm} $\mathbf{end}$
\State \hspace{0.1cm} $\{t\} \leftarrow$ GenerateTests($\mathcal{R}$, 1); \Comment{Generate one test input}
\State \hspace{0.1cm} $\mathit{DS}$ $\leftarrow$ $\mathit{DS}$ $\cup$ Execute($S$, $\{t\}$, $F$); \Comment{ Compute fitness for $t$ and add to $\mathit{DS}$}
\State \hspace{-0.1cm} $\mathbf{end}$ 
\State \hspace{-0.1cm} \textbf{return} $\mathit{DS}$
\end{algorithmic}

\end{algorithm}

\textbf{Logistic Regression Guided Test Generation. } Similar to Algorithm 4, Algorithm 5 uses the dataset $\mathit{DS}$ from  the preprocessing phase to train a logistic regression model  (lines~1-3). Since logistic
regression is a classification technique,  the quantitative labels in $\mathit{DS}$ are replaced with pass/fail labels before
training. A linear logistic regression model is represented as \hbox{$\log(\frac{p}{1-p}) = c + \sum_{i = 1}^{n}c_iv_i$} where $v_1, \ldots, v_n$ are the input variables,  $c_i$'s and $c$ are co-efficients, 
and $p$ is the probability of the pass class~\cite{logreg, dataminingbook}.  The algorithm then randomly samples a few test inputs in the search space and picks the one  closest to the logistic regression formula obtained by setting $p$ to  the percentage of the pass labels in $\mathit{DS}$ (line~4-5).
 

By assigning a value to $p$ in the logistic formula above, the formula turns into a linear equation. Figure~\ref{fig:logregexample} shows examples of such linear equations for different values of $p$ assuming that we have two variables $v_1$ and $v_2$. The line with $p = 0.9$ identifies a region where the majority of test inputs pass. On the other end of the spectrum, the line with $p = 0.3$ identifies a region where the majority of test inputs fail. Setting $p$ to the percentage of pass in $\mathit{DS}$ is a heuristic to identify a region that includes a mix of pass and fail test inputs. Our sampling should, therefore,  exploit this region. The test input selected on line~5 along with its fitness measure computed using $S$ is added to $\mathit{DS}$ (line~6).  The algorithm re-trains the logistic regression model whenever a test input  has been added to $\mathit{DS}$ (line~3). The algorithm returns the final dataset when the execution budget runs out (line~8). The execution budget expires when either the system has been executed to the point where the size of $DS$ reaches its desired limit, or our time budget is exhausted, depending on which occurs first.


\begin{figure}[t]
\includegraphics[width=0.4\columnwidth]{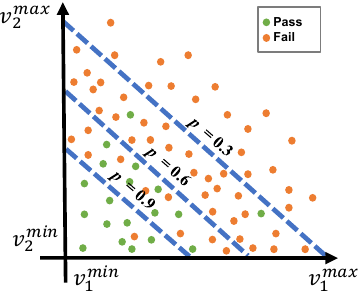}
\caption{Illustrating logistic regression lines for different values of $p$ assuming that we have two variables $v_1$ and $v_2$. The value of $p$ is used on line~5 of Algorithm~\ref{alg:alg5} to generate a test close to the regression border. }
\label{fig:logregexample}
\end{figure}

Similar to Algorithm 2, the hyperparameters of the regression-tree and logistic-regression models are tuned using Bayesian optimization the first time the models are built, and the same tuned hyperparameters are used in all future iterations.




\begin{algorithm}[t]
\caption{ML-guided test generation with logistic regression}
\label{alg:alg5}
\begin{flushleft}
\textbf{Input} $S$: System\\
\textbf{Input} $\mathcal{R} = \{R_1, \ldots, R_n\}$: Ranges $\mbox{for}$ input variables $v_1$ to $v_n$\\
\textbf{Input} $F$: Fitness $\mbox{Function}$\\
\textbf{Input} $\textsc{InitialDatasetSize}$: size of initial dataset\\
\textbf{Output} $\mathit{DS}$: A dataset to train failure models \\
\end{flushleft}
\begin{algorithmic}[1]    
\State \hspace{-0.1cm} $\mathit{DS}\leftarrow $  Preprocessing($S$, $\mathcal{R}$, $F$, $\textsc{InitialDatasetSize}$);\Comment{Run Algorithm~$1$ to generate an initial dataset}
\State \hspace{-0.1cm} \textbf{while} 
(execution budget remains) \textbf{do} 
\State \hspace{0.1cm} $\mathit{LogReg}$ $\leftarrow$ Train($\mathit{DS}$);
\State \hspace{0.1cm} $p \leftarrow $ Probability($\mathit{DS}$); \Comment{ probability of pass in $\mathit{DS}$}
\State \hspace{0.1cm} $t$ $\leftarrow$ GenerateCloseToRegBorder($\mathcal{R}$, $\mathit{LogReg}$, $p$); \Comment{ Select a test close to the regression border for $p$}
\State \hspace{0.1cm} $\mathit{DS}$ $\leftarrow$ $\mathit{DS}$ $\cup$ Execute($S$, $\{t\}$, $F$);\Comment{ Compute fitness for $t$ and add to $\mathit{DS}$}
\State \hspace{-0.1cm} $\mathbf{end}$ 
\State \hspace{-0.1cm} \textbf{return} $\mathit{DS}$
\end{algorithmic}

\end{algorithm}

\subsection{Building Failure Models}
\label{subsec:fms}
The output of the main loop in Figure~\ref{fig:fig1} is a set $\mathit{DS}$ of tuples $\langle t, F(t)\rangle$ where $t$ is a test input and $F(t)$ is its fitness value.  We first convert $\mathit{DS}$ into a dataset where test inputs are labelled by \emph{pass} and \emph{fail} labels. Provided with a labelled dataset, we use decision-rule models built using  RIPPER~\cite{ripper} to train failure models.  Decision-rule models generate a set of \element{IF-condition-THEN-prediction} rules where the \element{condition} is a conjunction of predicates over the input features and the \element{prediction} is either  pass or fail. In Section~\ref{sec:motivation}, we already showed two examples of such rules for spurious failures. When no domain knowledge is available,  one can directly use the input variables of the system ($S$) as features for learning. When domain knowledge is available, feature design for  decision-rule models can be improved in two ways: (1)~Excluding input variables that are orthogonal to the requirement under analysis.  For example, the prerequisite  for  the requirement $\varphi$ in Section~\ref{sec:motivation} is that the autopilot should be enabled, i.e., \element{APEng = on}.  As far as test generation for $\varphi$ is concerned, we need to set \element{APEng = on}, since otherwise, $\varphi$ holds vacuously. We thus do not use \element{APEng} as an input feature. For another example, in $\varphi$, we do not use the desired altitude as an input feature either, since the system is expected to satisfy $\varphi$ for any desired  altitude in the default range. (2)~Using domain knowledge to formulate features over multiple input variables. For NTSS, as discussed in Section~\ref{sec:motivation}, the goal is to identify limits on the traffic that can flow through NTSS classes  without compromising network quality. Based on domain knowledge, we know that flows have a cumulative nature. Hence, for NTSS, we use as features \emph{sums of subsets} of flow variables. Naturally, like in any feature engineering problem, one can hypothesize alternative ways of formulating the features and empirically determine the formulation leading to highest accuracy~\cite{machinelearningyearning}. 





\section{Evaluation}
\label{sec:evaluation}

In this section, we  evaluate our approach by answering the following research questions (RQs):

\textbf{RQ1 (Configuration).} \emph{Which surrogate-assisted technique offers the best trade-off between accuracy and efficiency?} We compare \emph{eight} surrogate-assisted  algorithms. These eight algorithms are: (a)~Algorithm 2 used with the seven  surrogate models in Table~\ref{tab:surrogatemodels} individually, and (b)~the dynamic surrogate algorithm (Algorithm 3) that uses the seven surrogate models simultaneously and selects the best model dynamically.  To measure accuracy, we check the correctness of the labels of the tests in the generated datasets; and, to measure efficiency, we evaluate the size of the generated datasets. We use the optimal algorithm for answering RQ2 to  RQ4.

\textbf{RQ2 (Effectiveness).} \emph{How accurate are the failure models generated by the surrogate-assisted and ML-guided  techniques?} We evaluate and compare the accuracy of the failure models obtained by surrogate-assisted and ML-guided algorithms as well as  those obtained based on randomly  generated test inputs (random baseline). 

\textbf{RQ3 (SoTA Comparison).} \emph{How accurate are the failure models generated in RQ2 compared to those generated by the state of the art (SoTA)?} We use the top-performing  technique from RQ2 to compare against SoTA. 
Among the existing approaches that build failure-inducing models~\cite{kampmann2020does, gopinath2020abstracting, kapugama2022human},  we select the Alhazen framework~\cite{kampmann2020does}, since it uses interpretable machine learning. 
While Alhazen is geared towards systems with structured inputs (as opposed to systems with numeric inputs, i.e., the focus of our work), in the absence of baselines for systems with numeric inputs, Alhazen  is our best baseline for comparison. To be able to compare with Alhazen, we adapt it to numeric-input systems, as we describe in Section~\ref{subsec:rq3}.



\textbf{RQ4 (Usefulness).} \emph{How useful are  failure models for identifying spurious failures?} We answer this question for the most accurate failure models from RQ2 and for the two CI systems, namely NTSS and autopilot, discussed in Section~\ref{sec:motivation}. NTSS and autopilot are representative examples of industrial systems in the network and CPS domains, respectively. For both systems, we validate the failure-inducing rules against domain knowledge to determine whether the resulting failures are genuinely spurious. 

\subsection{Study Subjects} 
\label{subsec:study}

Our study subjects, which are listed in Table~\ref{tab:simulinkmodels}, originate  from the network and CPS domains. Below, we introduce our study subjects and discuss how these subjects satisfy assumptions \textbf{A1} and \textbf{A2} provided at beginning of Section~\ref{sec:approach}. 


\begin{table*}[t]
\caption{Names, descriptions, the number of requirements and the identifiers of our study subjects. For each subject, we indicate  if it is computer-intensive (CI). All artifacts including requirements statements are available in our supplementary material~\cite{requirements}. }
\label{tab:simulinkmodels}
\scalebox{0.8}{
\begin{tabular}{|l|l|l|l|l|l|}
\hline
\multicolumn{1}{|c|}{\textbf{Name}} & \multicolumn{1}{p{3cm}|}{\textbf{Description}}                                                    & \multicolumn{1}{c|}{\textbf{\#Reqs}} & {\textbf{ID}} & \multicolumn{1}{c|}{\textbf{CI}} \\ \hline
 Tustin                             & \multicolumn{1}{p{7cm}|}{ A common flight control utility for computing the Tustin Integration -- A Simulink model with                          57 blocks.}                              & 9                                   &TU1$\ldots$TU9& \ding{55}                                                               \\ \hline
 Regulator                          & \multicolumn{1}{p{7cm}|}{A regulators inner loop architecture used in many feedback control applications -- A Simulink model with 308  blocks.}                                 &      1          & REG  &\ding{55}                                   \\ \hline
                                Nonlinear Guidance                 & \multicolumn{1}{p{7cm}|}{A nonlinear algorithm for generating a guidance command for an air vehicle -- A Simulink model with                    373   blocks.}                                 &   1                                 & NLG & \ding{55}  
\\ \hline
                                Finite State Machine                 & \multicolumn{1}{p{7cm}|}{A finite state machine to enable
autopilot mode if a hazardous situation is identified -- A Simulink model with                       303   blocks.}                                 &   1                                  & FSM & \ding{55}    
\\ \hline
                                 Autopilot                          & \multicolumn{1}{p{7cm}|}{A single-engine, high-wing, propeller-driven aircraft simulation with all six degrees of freedom -- A Simulink model with  1549 blocks.}                                 & 3                                     &  AP1, AP2, AP3 &  \ding{51}                                \\ \hline

 Network Traffic Shaping System & \multicolumn{1}{p{7cm}|}{An NTSS testbed~\cite{enrich} developed using three virtual machines and based on OpenWRT~\cite{openwrt}.} 

 & 1 & NTSS & \ding{51}  \\ \hline

\end{tabular}}
\vspace*{-.35cm}
\end{table*}
\emph{Network-system subject.} Our network-system subject is  the Network Traffic Shaping System (NTSS) discussed in Section~\ref{sec:motivation}. To test NTSS, we transmit flows with different bandwidth values into different NTSS classes. A test input for NTSS is  defined as a tuple $t = (v_1, \ldots, v_n)$ where $n$ is the number of NTSS classes, and each variable $v_i$ represents the bandwidth of the data flow going through class $i$. The fitness function for NTSS measures the network quality based on the well-known mean opinion score (MOS) metric~\cite{mos}. This fitness function ensures assumption \textbf{A2}~\cite{enrich}.  For our experiments, we use an NTSS setup based on an industrial small-office and home-office use case from our earlier work~\cite{enrich}.  This setup runs Common Applications Kept Enhanced (CAKE)~\cite{cakepaper}, which is an advanced and widely used traffic-shaping algorithm. The setup uses the 8-tier mode of CAKE known as diffserv8~\cite{cakepaper, CAKE}, i.e., the number of NTSS classes is $8$.

\emph{Simulink subjects.}  Simulink~\cite{chaturvedi2017modeling} is a widely used language for specification and simulation  of CPS.  The inputs and outputs of a Simulink model are represented using signals. A typical input-signal generator for Simulink characterizes each input signal using a triple $(\mathit{int}, R, n)$ such that $\mathit{int}$ is an interpolation function,  $R$ is a value range, and $n$ is a number of control points~\cite{tuncali2019requirements,arrieta2019pareto}. 
Let $(x, y)$ be a control point. The value of $x$ is from the signal's time domain, and the value of $y$ is from range $R$. Provided with $n$ control points, 
the interpolation function $R$ (e.g., piecewise constant, linear or piecewise cubic) constructs a signal by connecting the control points~\cite{tuncali2019requirements}. It is usually assumed that the input signals for a Simulink model have the same time domain. Further, for the purpose of testing, we make the common assumption that the control points are equally spaced over the time domain, i.e., the control points are positioned at a fixed time distance~\cite{tuncali2019requirements,arrieta2019pareto,miningassumption}. Hence, to generate test inputs, we only need to vary the $y$ variable of control points in the range $R$.  Note that the type of the interpolation function is fixed for each Simulink-model input as the interpolation function type is determined by the meaning of the input signal. For example,  a reference signal is often a constant or step function. As a result,  we can exclude the interpolation function from the  test-input signals, and define each test-input signal as a vector of $n$ control variables taking their values from $R$.  Simulink models often have clearly stated requirements. To define a fitness function for each requirement, we  encode the requirement in RFOL -- a variant of the signal temporal logic which is expressive enough for capturing many CPS requirements~\cite{menghi2019generating}. For our fitness function, we utilize the RFOL semantic function -- a quantitative measure that conforms to assumption \textbf{A2} as shown by Menghi et al.~\cite{menghi2019generating}.


In our evaluation, we use a public-domain benchmark of Simulink specifications from Lockheed Martin~\cite{lockheedmartin}. This benchmark provides a set of representative CPS systems that are shared by Lockheed as verification-and-validation challenge subjects for researchers and quality-assurance tool vendors.  The benchmark is comprised of eleven specifications, of which six were not useful for our evaluation. These six specifications had requirements that either did not fail, or failed for all test inputs and thus, their failure models could be trivially defined as the entire input space. In our experiments,  we focus on five of the Simulink specifications from Lockheed's benchmark. The first five rows of Table~\ref{tab:simulinkmodels} list these specifications that have a total of $15$ requirements combined. 



In total, we have $16$ requirements: one for NTSS, and $15$ for the five Simulink specifications listed in Table~\ref{tab:simulinkmodels}. As discussed in Section~\ref{sec:approach}, we define one fitness function per requirement.  Hence, in total, we have $16$ different subjects for our evaluation. Among these, four are compute-intensive (CI) and twelve are non-CI. Both NTSS and autopilot are CI: On average, each execution of NTSS takes $\approx 4.5$min, and each execution of autopilot takes $\approx 0.5$min. The execution times for non-CI subjects are negligible ($<1s$).  All experiments were conducted on a machine with a 2.5 GHz Intel Core i9 CPU and 64~GB of DDR4 memory.

\subsection{RQ1-Configuration} 
\label{subsec:rq1}
We compare eight versions of the surrogate-assisted algorithm. Seven are Algorithm~2 used with an individual surrogate model from those in Table~\ref{tab:surrogatemodels}. We refer to each of these algorithms as \element{SA-XX} where \element{XX} is the name of the surrogate model from Table~\ref{tab:surrogatemodels}. For example, \element{SA-NN} refers to Algorithm~2 used with NN. The final (i.e., eighth) algorithm is the dynamic surrogate-assisted one (Algorithm~3). We refer to Algorithm~3 as \element{SA-DYN}.  

For RQ1, we use the $12$ non-CI  subjects in Table~\ref{tab:simulinkmodels}. Performing RQ1 experiments on CI subjects would be prohibitively expensive. For example, an approximate estimate for the execution time of the experiments required to answer RQ1 is over a year, if the experiments are performed on NTSS using the same experimentation platform. Thus, we opt for the non-CI subjects for RQ1.




\emph{Setting.} For each  subject, we run the eight \element{SA-XX} algorithms for an equal time budget. The time budget given for each subject depends on the subject execution time. The detailed time budgets are available in our supplementary material~\cite{table4}.  To account for randomness, we repeat each algorithm for each subject ten times.

\emph{Metrics.} Recall that the output of the main loop in Figure~\ref{fig:fig1}  for surrogate-assisted algorithms is a set $\mathit{DS}$  where the test inputs are labelled with either predicted or actual fitness values. To measure  efficiency, we take the cardinality of the generated dataset, $\mathit{DS}$. Since the algorithms have the same time budget, an algorithm is more efficient than another if it generates a larger dataset.  To construct failure models, we classify test inputs as pass or fail based on their fitness values. A dataset is accurate if it contains few test inputs with \emph{incorrect labels}, i.e.,  test inputs with inconsistent  pass/fail labels based on their predicted versus actual fitness values. To obtain the actual fitness values for all test inputs, we run the system using those test inputs for which only surrogate-assisted algorithms provided predicted fitness values. Note that this step is intended exclusively for our experiments and is not a component of our main approach. Next, we count the number of tests in which the predicted label differs from the actual one.
The number of incorrectly labelled tests serves as a measure of error or inaccuracy for the surrogate-assisted algorithms. 


\begin{figure}[t]
    \centering
    \includegraphics[width=1\columnwidth]{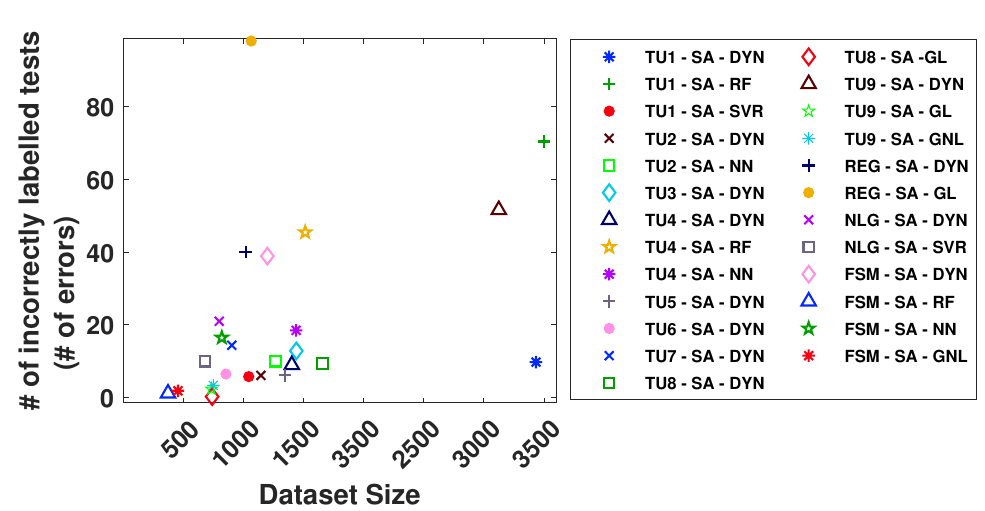}
    
    \caption{Comparing datasets generated by eight different surrogate-assisted algorithms with respect to the number of errors in the datasets and the dataset size.}\label{fig:rq1-1}
    
\end{figure}

\begin{figure}[t]
    \centering
    \includegraphics[width=0.7\columnwidth]{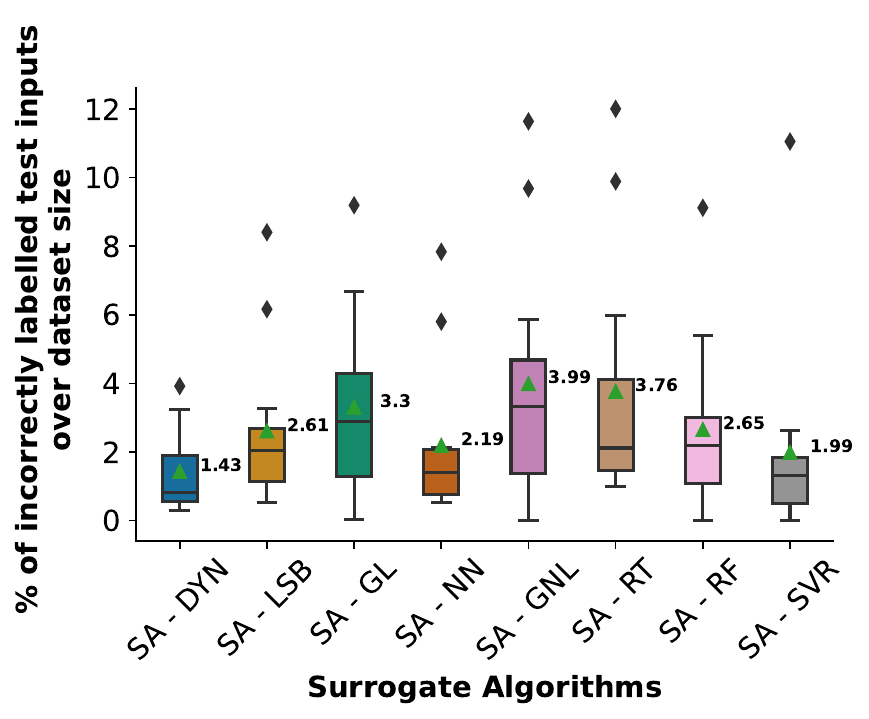}
    \vspace*{-.3cm}
    \caption{Percentages of the incorrectly labelled tests over the dataset size for different surrogate-assisted algorithms.}\label{fig:rq1-2}
    \vspace*{-.5cm}
\end{figure}

\emph{Results.} The scatter plot in Figure~\ref{fig:rq1-1} shows the results of the experiments for RQ1. The x-axis indicates  $|\mathit{DS}|$ and the y-axis indicates  the number of incorrectly labelled tests in $\mathit{DS}$. Each point shows the result of applying one \element{SA-XX} algorithm to one subject. The $12$ subjects are indicated by \element{TU1} to \element{TU9}, \element{REG}, \element{NLG}, and \element{FSM} (see Table~\ref{tab:simulinkmodels}). For each subject, an algorithm is considered better when it generates larger datasets with fewer errors. Since we compare eight algorithms for $12$ subjects, we would need $96$ points to show all the results. To reduce clutter, for each subject, we only show the Pareto Front (PF) points. That is, for each subject, we only show the algorithms that are not dominated by other algorithms either in terms of the number of errors or in terms of the dataset size. For example, for subject \element{TU1}, algorithms \element{SA-DYN}, \element{SA-RF} and \element{SA-SVR} dominate other algorithms and offer the best trade-off between the number of errors and the dataset size. As Figure~\ref{fig:rq1-1} shows, for four subjects, \element{SA-DYN} is the only best trade-off (i.e., PF point), and for eight subjects, it is one of the best PF points. For the latter eight cases, 
\element{SA-DYN} offers an alternative that, compared to the other PF points, either has  considerably fewer errors  while its dataset is not much smaller, or its dataset is considerably larger  while the number of errors is not much higher. For example, for \element{TU1}, \element{SA-DYN},  compared to \element{SA-RF},  provides $60$ less incorrectly labelled tests, while the dataset sizes are almost the same ($3433$ for \element{SA-DYN} vs. $3500$ for \element{SA-RF}). Also, compared to \element{SA-SVR}, \element{SA-DYN} provides a larger dataset ($3433$ vs. $1045$) while the number of incorrectly labelled tests is almost the same ($10$  vs. $6$). 

Figure~\ref{fig:rq1-2} shows the ratios of  the number of errors (i.e., the number of incorrectly labelled tests) over $|\mathit{DS}|$ for different SA algorithms and for all the $12$ subjects. The \element{SA-DYN} algorithm has the lowest average error which is $28$\% lower than that of the second best algorithm, \element{SA-SVR}, i.e., $\frac{1.99-1.43}{1.99} = 28$\%.  We compare the results in Figure~\ref{fig:rq1-2}  using the non-parametric pairwise Wilcoxon rank-sum test~\cite{wilcoxon} and the Vargha-Delaney's $\hat{A}_{12}$ effect size~\cite{vargha2000critique}. The \element{SA-DYN} algorithm is statistically significantly better than other algorithms with a high effect size for \element{GL}, \element{GNL} and \element{LSB}, a medium effect size for \element{RT},  a small effect size for \element{NN} and \element{RF}, and a negligible effect size for \element{SVR}. The comparison of the dataset sizes shows that \element{SA-DYN} generates datasets that are  significantly larger than those generated by \element{SVR} with a large effect size. Further,  \element{SA-DYN} generates significantly larger datasets that are at least $33$\% larger than those obtained from other algorithms. Figures comparing dataset sizes and statistical tests for RQ1 are available \hbox{in our supplementary material~\cite{figure9, table5}.} 

\begin{framed}
Our evaluation for RQ1 performed using seven surrogate-model types in the literature (see Table~\ref{tab:surrogatemodels}) shows that, compared
to using surrogate models individually, our dynamic surrogate-assisted approach provides the best trade-off between  accuracy
and  efficiency. 
\end{framed}

\subsection{RQ2-Effectiveness} 
\label{subsec:rq2}
We compare \element{SA-DYN}, i.e.,  the best approach identified in RQ1, with the two ML-guided algorithms described in Section~\ref{subsec:mlguided} as well as a standard adaptive random-search algorithm. In the remainder of this section, we refer to \element{SA-DYN} as \element{SA}. We use \element{RT} and \element{LR} to refer to the ML-guided algorithms that employ regression tree (Algorithm~4) and  logistic regression (Algorithm~5), respectively. We use \element{RS} for adaptive random search.


 

\emph{Setting.} We apply the four algorithms (i.e., \element{SA}, \element{RT}, \element{LR}, and \element{RS}) to the 16 subjects in Table~\ref{tab:simulinkmodels}. For each CI subject, we execute the four algorithms for an equal time budget and then compare the results. For the non-CI subjects, however, fixing the time budget  may favour \element{RS}, as the other three algorithms have an additional overhead for training ML models.  This overhead time is negligible when compared to the execution time for CI subjects. But, for non-CI subjects, we can execute many tests within the overhead time, which can skew the results in favour of \element{RS}.  Therefore, to ensure that the results are valid for CI subjects, we follow the  approach proposed by Menghi et al.~\cite{menghi2020approximation}. Specifically, we use the execution time of CI subjects to limit the number of test inputs that each algorithm executes for non-CI objects. Briefly, to compare two algorithms with different overhead times, we allow the algorithm with the lower overhead time to execute $x$ additional test inputs such that $x$ multiplied by the execution time of a typical CI subject (instead of a non-CI subject) is equal to the difference in the two algorithms' overhead time. For the non-CI Simulink subjects in Table~\ref{tab:simulinkmodels}, we use the average execution time of the Simulink CI subject, i.e., autopilot. Given a time budget, we compute the maximum number of test executions that each of the \element{SA}, \element{LR}, \element{RT}, and \element{RS} algorithms can perform within this time budget for autopilot. We then use these numbers to cap the number of test executions for each algorithm when we compare them for the non-CI models in Table~\ref{tab:simulinkmodels}. The time budget we consider for comparing these algorithms for CI subjects and the maximum number of test executions we use to compare them for non-CI subjects are available in our supplementary material~\cite{table6, table7}. We repeat each algorithm twenty times for each subject except NTSS. For NTSS, due to its large execution time, we repeat each algorithm only ten times.




\emph{Metrics.}  We use the datasets created by \element{SA}, \element{LR}, \element{RT}, and \element{RS} to build decision rule models (DRM). For hyperparameter tuning, we use Bayesian optimization~\cite{bayesian}. To avoid bias towards any particular algorithm, we use for tuning the union of the datasets obtained from our four algorithms. Having fixed the hyperparameters, we  train a DRM separately for each dataset obtained from each repetition of our four algorithms. We evaluate the DRMs using three metrics: \emph{accuracy}, \emph{precision} and \emph{recall}.  Accuracy is the number of correctly predicted tests over the total number of tests. Since DRMs are mainly used to predict the failure class, we compute precision and recall for the failure class as follows: Precision is the number of fail-class predictions that actually belong to the fail class, and recall is the number of fail-class predictions out of all the actual failed tests in the dataset. We use randomly generated test inputs within the variables' default ranges to measure the accuracy, precision and recall of each DRM. 

\emph{Features for learning.} For the Simulink subjects, we use as features the individual input variables of each subject model but exclude the following two kinds of variables: (1)~Variables that are explicitly fixed to a value in a requirement (e.g., variable \element{APEng} discussed in Section~\ref{subsec:fms}). (2)~Reference variables that indicate the desired value of a controlled process, noting that the system is expected to satisfy its requirements for \emph{any} value in a reference variable's valid range. As such, reference variables cannot contribute to creating failure conditions. The desired altitude variable discussed in Section~\ref{subsec:fms} is an example of a reference variable. 

For NTSS, as discussed in Section~\ref{subsec:fms}, we consider alternative features as follows: the set of all individual variables (i.e., individual NTSS classes), sums of two variables, sums of three variables, \ldots, and the sum of all eight variables.  We then create, for each input feature, one DRM  based on a dataset obtained by each of the four algorithms. In total, for each algorithm, we create $248$ DRMs for NTSS. That is, the sets of all subsets larger than two ($247$) and the set of all individual variables.  Given the large number of hypothesized input features, we evaluate the accuracy of the resulting DRMs and keep the  input features that yield reasonably high accuracy over multiple runs of \element{SA}, \element{LR}, \element{RT}, \element{RS}. This results in the retention of  two input features for NTSS with an accuracy higher than $80$\%. 



\emph{Results.}
Figures~\ref{fig:rq2-1}(a)-(c) compare across all the $16$ subjects  the average accuracy of DRMs obtained by \element{SA}, \element{LR} and \element{RT} (on y-axis) against the average accuracy of DRMs obtained by \element{RS} (on x-axis). Each point in each of Figures~\ref{fig:rq2-1}(a)-(c) corresponds to one study subject.  A blue point indicates that the difference in accuracy is statistically significant as per  the Wilcoxon rank-sum test. The DRMs obtained using \element{SA} are significantly more accurate than those obtained using \element{RS} for $14$ out of $16$ subjects, including  all the CI subjects. The accuracy of the DRMs  obtained using \element{LR} is significantly better than that obtained using \element{RS} for seven subjects. 
The accuracy of the DRMs  obtained using \element{RT} is significantly better than that obtained using \element{RS} for nine subjects.  Overall for all the subjects, \element{SA} has the highest average accuracy ($83\%$), followed by \element{RT} and \element{LR} with average accuracies of $78\%$ and $76\%$, respectively.  \element{RS} has the lowest average accuracy ($72\%$). Finally, the accuracy of \element{SA}, \element{RT} and \element{LR} is significantly better than that of \element{RS}. The effect size for \element{SA} versus \element{RS} is medium, and the effect size for  \element{RT} and  \element{LR} versus  \element{RS} is small. 

\begin{figure}[t]
    \centering
    \includegraphics[width=0.7\textwidth]{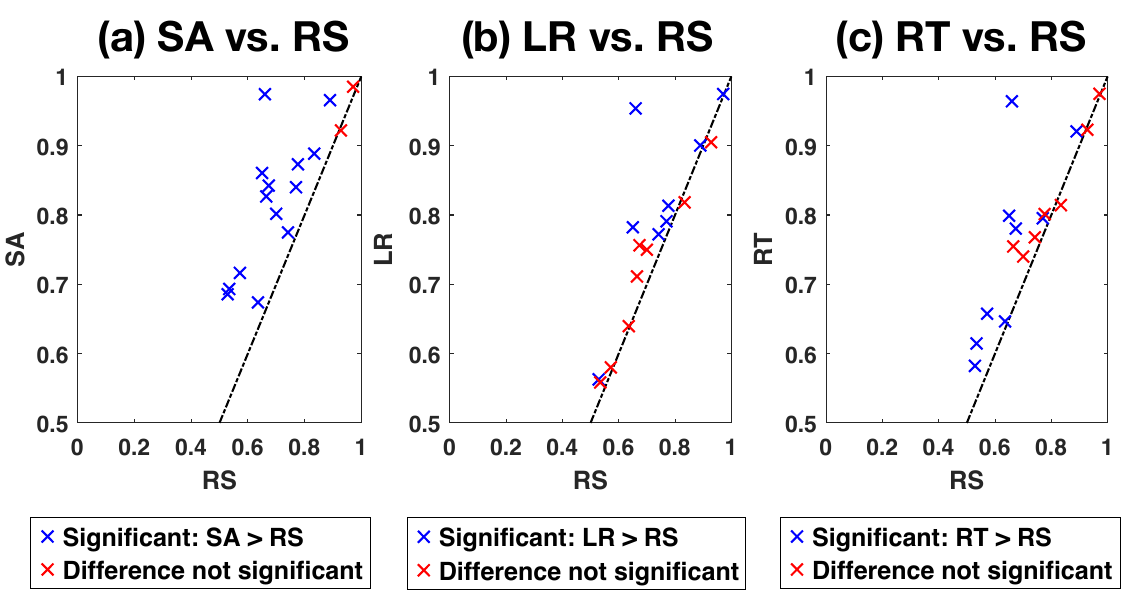} 
    
    \caption{Comparing the accuracy of decision-rule models obtained based on the datasets generated by \element{SA}, \element{LR}, and \element{RT} against those obtained by \element{RS} for all the $16$ subjects in Table~\ref{tab:simulinkmodels}.}\label{fig:rq2-1}
    
\end{figure}

Figure~\ref{fig:avgaccbudget} compares the average accuracy of the DRMs obtained using \element{SA}, \element{LR}, \element{RT} and \element{RS} for our $16$ subjects as the  execution-time budget is varied from $50$\% to $100$\%.  Note that in our experiments, we dedicate $50$\% of the time budget to the prepossessing step. Therefore, Figure~\ref{fig:avgaccbudget} compares the impact of the four algorithms over the remaining $50$\% of the execution-time budget. As the figure shows, \element{SA} consistently has the highest average accuracy. Further, the average accuracy of \element{RS} reaches a plateau, whereas the other three algorithms keep improving as the budget increases. The main reason for this difference is that \element{RS}, unlike the other algorithms, does not use machine learning models to guide its test input sampling.  Specifically, \element{RS} generates test inputs randomly across the entire input space. In contrast, \element{RT} and \element{LR} exploit boundary regions that separate passing and failing test inputs, resulting in a steady increase in accuracy as the search budget grows.  Further, \element{SA} utilizes surrogate models, generating significantly more test inputs compared to the other algorithms within the same allotted time budget. As a result, SA achieves the highest average accuracy among all the algorithms.


    
    

\begin{figure}[t]
    \centering
    \includegraphics[width=.5\columnwidth]{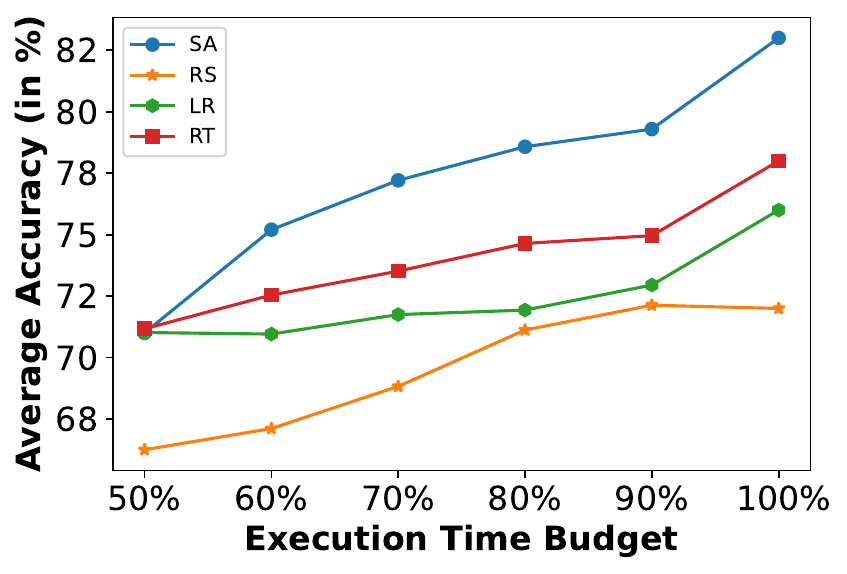}
     
    \caption{Average accuracy of the DRMs obtained using  \element{SA}, \element{RL}, \element{RT} and \element{RS}  for  the $16$ subjects as the time budget is varied.}\label{fig:avgaccbudget}
   
\end{figure}

Figure~\ref{fig:rq2-2} compares the recall and precision for all the DRMs obtained using \element{SA}, \element{LR}, \element{RT} and \element{RS} for our $16$ subjects. Recall measures the ability of DRMs to precisely identify the failure conditions, whereas precision assesses the ability of DRMs to generate failure instances correctly. The \element{SA} algorithm has the highest average recall ($88\%$), followed by \element{LR} ($87\%$) and \element{RT} ($85\%$). \element{RS} has the lowest average recall ($83\%$). Moreover, \element{SA} has the highest average precision ($72\%$), followed by \element{RT} ($64\%$) and \element{LR} ($62\%$) and \element{RS} has the lowest average precision ($57\%$). The recall of \element{SA}, \element{RT} and \element{LR} is significantly better than that of \element{RS} with a medium effect size for \element{SA}, a small effect size for \element{LR}, and negligible effect size for \element{RT}. Likewise, the precision of \element{SA}, \element{RT} and \element{LR} is significantly better than that of \element{RS} with a medium effect size for \element{SA}, small effect size for \element{RT} and negligible effect size for \element{LR}. Finally, the accuracy, recall and precision values for \element{SA} are significantly higher than those for \element{RT} and \element{LR}. Precision and recall for \element{SA} exhibit similar trends to that for accuracy (Figure~\ref{fig:avgaccbudget}). Detailed charts  comparing precision and recall for the four algorithms as the execution-time budget  varies are available online~\cite{figures16to21,tables8to13,tables14to19}.

\begin{figure}[t]
    \centering
    \includegraphics[width=0.88\columnwidth]{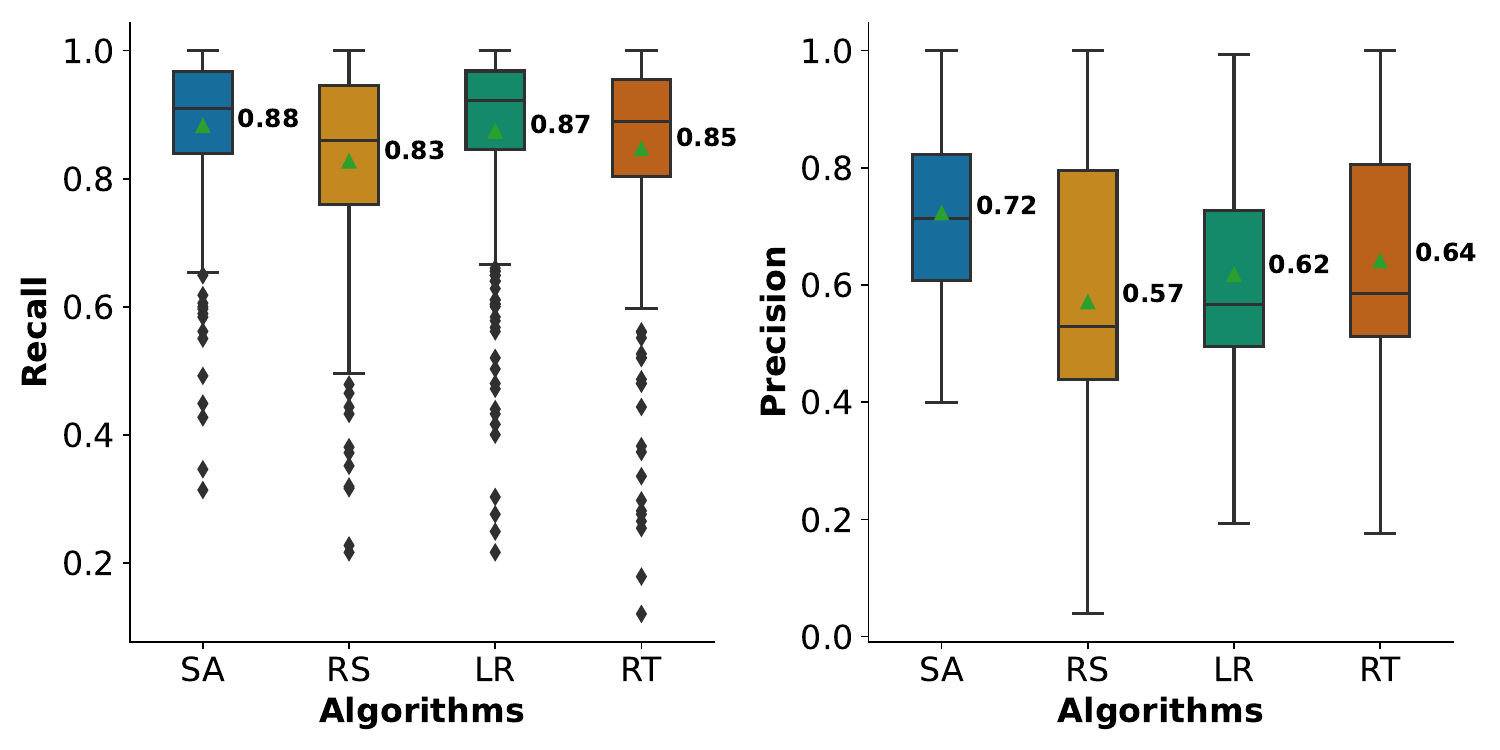}
    \vspace*{-.3cm}
    \caption{Recall and Precision  for the DRMs obtained based on  \element{SA}, \element{RL}, \element{RT} and \element{RS} for all the $16$ subjects in Table~\ref{tab:simulinkmodels}.}\label{fig:rq2-2}
    \vspace*{-.25cm}
\end{figure}

\begin{framed}
Our evaluation for RQ2 performed over all our study subjects shows that our dynamic surrogate-assisted approach yields failure models with significantly higher accuracy, precision and recall compared to those obtained using  ML-guided algorithms and a random baseline. 
\end{framed}


\subsection{RQ3-SOTA Comparison}
\label{subsec:rq3}
We compare \element{SA} -- the top-performing algorithm from RQ2 --  with an adaptation of Alhazen to numeric-input systems.  We hereafter use \element{SoTA} to refer to this adaptation, discussed next. Similar to \element{SA}, in \element{SoTA},  we generate test inputs  according to the description in Section~\ref{subsec:study}. The workflow of  \element{SoTA} then matches the steps in Figure~\ref{fig:fig1} with the  difference that the model trained and refined in the main loop (i.e. step~3) is always a decision tree model; this decision tree is returned as the failure model at the end. Recall that our approach has a separate step, i.e., step~5, after the main loop to build failure models from the data generated by different algorithms. This step is not required in \element{SoTA}, noting that the model that \element{SoTA} refines during its main loop is used as the failure model. Algorithm~$6$ shows our implementation of \element{SoTA}.  
 \element{SoTA} starts from an initial dataset (line~1). In each iteration, it builds a decision tree on the dataset (line~3).  It then generates test inputs using all the paths in the decision tree (lines~4-10). These test inputs are executed and added to the dataset along with their labels (line~11). The final decision tree is returned on line~14.

\begin{algorithm}[t]
\caption{Our implementation of SoTA}
\label{alg:alhazen}
\begin{flushleft}
    
\textbf{Input} $S$: System\\
\textbf{Input} $\mathcal{R} = \{R_1, \ldots, R_n\}$: Ranges $\mbox{for}$ Input variables $v_1$ to $v_n$\\
\textbf{Input} $F$: Fitness $\mbox{Function}$\\
\textbf{Input} $\textsc{InitialDatasetSize}$: size of initial dataset\\
\textbf{Output} $\mathit{DecisionTree}$: A decision tree model (failure model) \\
\end{flushleft}
\begin{algorithmic}[1]
\State \hspace{-0.1cm} $\mathit{DS}\leftarrow $  Preprocessing($S$, $\mathcal{R}$, $F$, $\textsc{InitialDatasetSize}$); \Comment{Run Algorithm~$1$ to generate an initial dataset}
\State \hspace{-0.1cm} \textbf{while} 
(execution budget remains) \textbf{do} 
\State \hspace{0.1cm} $\mathit{DecisionTree}$ $\leftarrow$ Train($\mathit{DS}$);
\State \hspace{0.1cm} Let $P_{1}$,..., $P_{q}$ be all the paths obtained from  $\mathit{DecisionTree}$; \Comment{ Extract all the paths from the decision tree}
\State \hspace{0.1cm} \textbf{for} each path $P_{k}$ s.t.  $k \in \{1, \ldots, q\}$ \textbf{do} \Comment{ Generate a test in each path based on the ranges obtained from that path}
\State \hspace{0.1cm} \hspace{0.1cm} Let $R'_{i_1}$,..., $R'_{i_m}$ be reduced ranges obtained from $P_{k}$;
\State \hspace{0.1cm} \hspace{0.1cm} \textbf{for} each variable $v_{i_j}$ s.t.  $j \in \{1, \ldots, m\}$ \textbf{do}
\State \hspace{0.1cm} \hspace{0.1cm} \hspace{0.1cm} $\mathcal{R} \leftarrow (\mathcal{R} \setminus \{R_{i_j}\}) \cup \{R'_{i_j}\}$; \Comment{ Replace the range $R_{i_j}$ of $v_{i_j}$ in   $\mathcal{R}$ with the new reduced range $R'_{i_j}$ from line 6}
\State \hspace{0.1cm} \hspace{0.1cm}  $\mathbf{end}$
\State \hspace{0.1cm} \hspace{0.1cm} $\{t\} \leftarrow$ GenerateTests($\mathcal{R}$, 1); \Comment{ Generate a test in path $P_k$}
\State \hspace{0.1cm} \hspace{0.1cm} $\mathit{DS}$ $\leftarrow$ $\mathit{DS}$ $\cup$ Execute($S$, $\{t\}$, $F$); \Comment{ Compute fitness for $t$ and add to $\mathit{DS}$}
\State \hspace{0.1cm} $\mathbf{end}$
\State \hspace{-0.1cm} $\mathbf{end}$ 
\State \hspace{-0.1cm} \textbf{return} $\mathit{DecisionTree}$
\end{algorithmic}
\end{algorithm}

\emph{Setting.} For this comparison, we apply \element{SoTA} to our CI-subjects in Table~\ref{tab:simulinkmodels}, i.e., NTSS, AP1, AP2 and AP3. For the decision tree parameters, e.g. maximum depth of tree and class weight, we use the same parameters as in the original study~\cite{kampmann2020does, alhazenparameters}. 
We execute \element{SoTA} for the same time budget as \element{SA}. For \element{SA}, we use the dataset generated in RQ2. We repeat \element{SoTA} for twenty times for each subject except for NTSS. For NTSS, we repeat it only ten times due to the expensive execution time. 


\emph{Metrics.} In order to compare \element{SoTA} and \element{SA},  we build decision trees based on the datasets generated by \element{SA} in RQ2. To do so, we use the same decision tree parameters as those used by \element{SoTA}. We compare the decision trees using the three metrics explained in RQ2, i.e. accuracy, precision for fail class and recall for fail class. We also use the same test set  utilized in RQ2.

\emph{Results.} Figure~\ref{fig:acc-sa-al} compares the average accuracy of the decision trees obtained by \element{SA} (on y-axis) against those obtained by \element{SoTA} (on x-axis) across the four CI subjects in Table~\ref{tab:simulinkmodels}. Similar to Figure~\ref{fig:rq2-1}, each point on Figure~\ref{fig:acc-sa-al} corresponds to one study subject and a blue point denotes a statistically significant difference in accuracy, determined using the Wilcoxon rank-sum test. As figure~\ref{fig:acc-sa-al} shows, the decision trees obtained using \element{SA} are significantly more accurate than those obtained by \element{SoTA}, for three out of the four subjects with a medium effect size. For the fourth subject there is no statistically significant difference. 

Figure~\ref{fig:acc-sa-al-overtime} compares the average accuracy of the decision trees obtained using  \element{SA} and those obtained by \element{SoTA} for the four CI subjects as the execution-time budget is varied from $50\%$ to $100\%$. As the figure shows, the average accuracy of \element{SA} is consistently higher than \element{SoTA} as the time budget increases. 

Finally, Figure~\ref{fig:precisionrecall-SAvsAL} compares the recall and precision for all the decision trees obtained using \element{SA} and those obtained by \element{SoTA} for the four subjects. As shown by the figure, the average recall of \element{SA} ($85\%$) is higher than \element{SoTA} ($83\%$). Further, the recall of \element{SA} is significantly better than \element{SoTA} with small effect size. Moreover, the average precision of \element{SA} ($76\%$) is higher than \element{SoTA} ($73\%$). Similar to recall, the precision of \element{SA} is significantly better than \element{SoTA} with small effect size. 
\begin{framed}
Our evaluation for RQ3 performed using dynamic surrogate-assisted and the state-of-the-art approach over CI subjects indicates that our dynamic surrogate-assisted approach yields failure models with higher accuracy, precision and recall compared to those obtained from the state-of-the-art approach. 
\end{framed}

\begin{figure}[t]
    \centering
    \includegraphics[width=0.3\columnwidth]{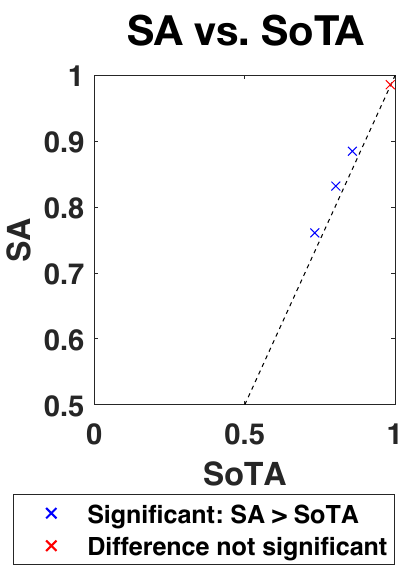}
    \caption{Comparing the accuracies of the decision trees obtained on the datasets generated by \element{SA} and the decision trees returned by \element{SoTA} for all the four CI subjects in Table~\ref{tab:simulinkmodels}.}\label{fig:acc-sa-al}
    \vspace*{-.25cm}
\end{figure}
\begin{figure}[t]
    \centering
    \includegraphics[width=0.5\columnwidth]{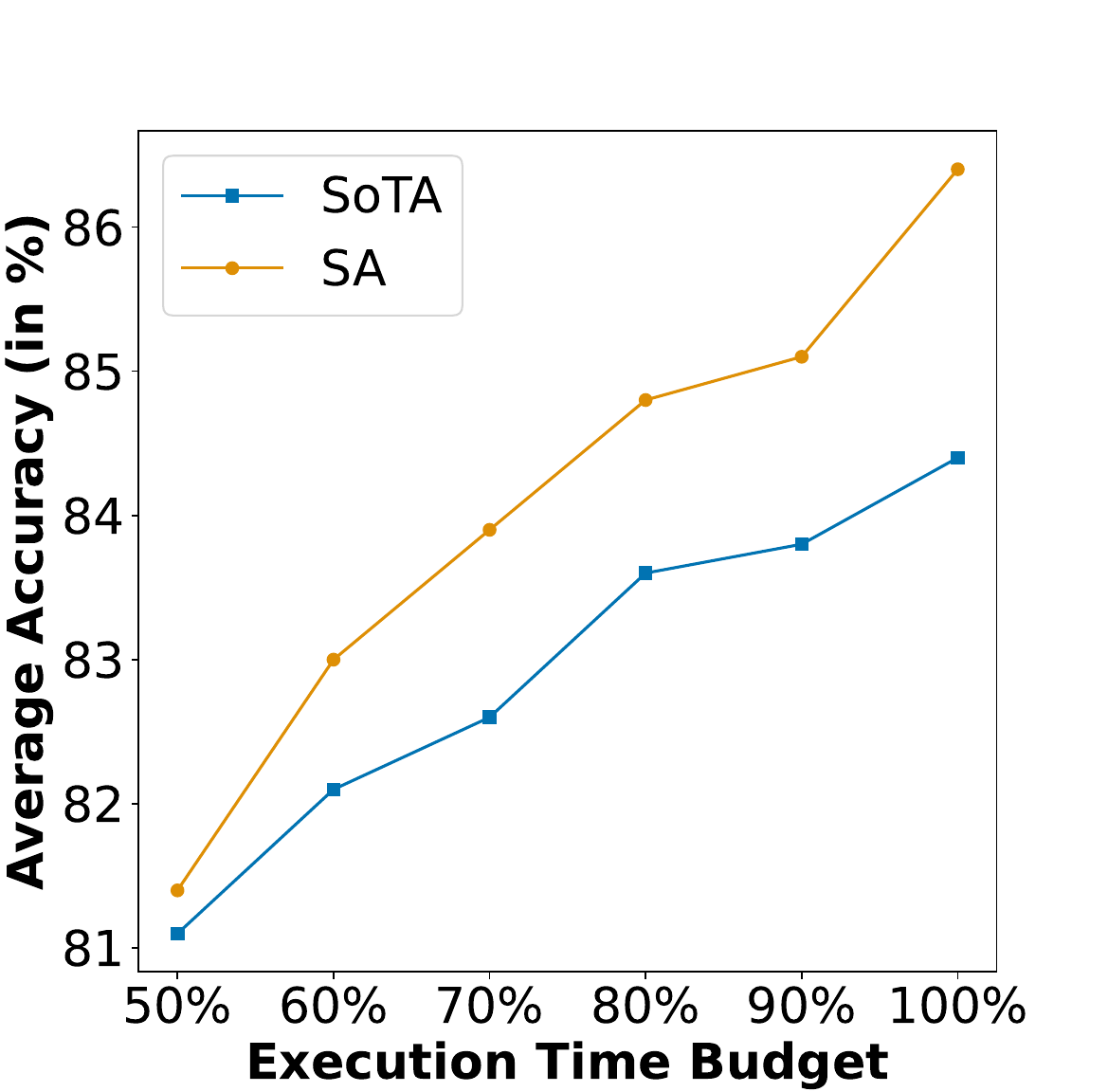}
    \vspace*{-.3cm}
    \caption{Average accuracy of the decision trees obtained using \element{SA} and \element{SoTA} for the four CI subjects as the time budget is varied.}\label{fig:acc-sa-al-overtime}
    \vspace*{-.25cm}
\end{figure}
\begin{figure}[t]
    \centering
    \subfloat{{\includegraphics[width=6cm, height = 6cm]{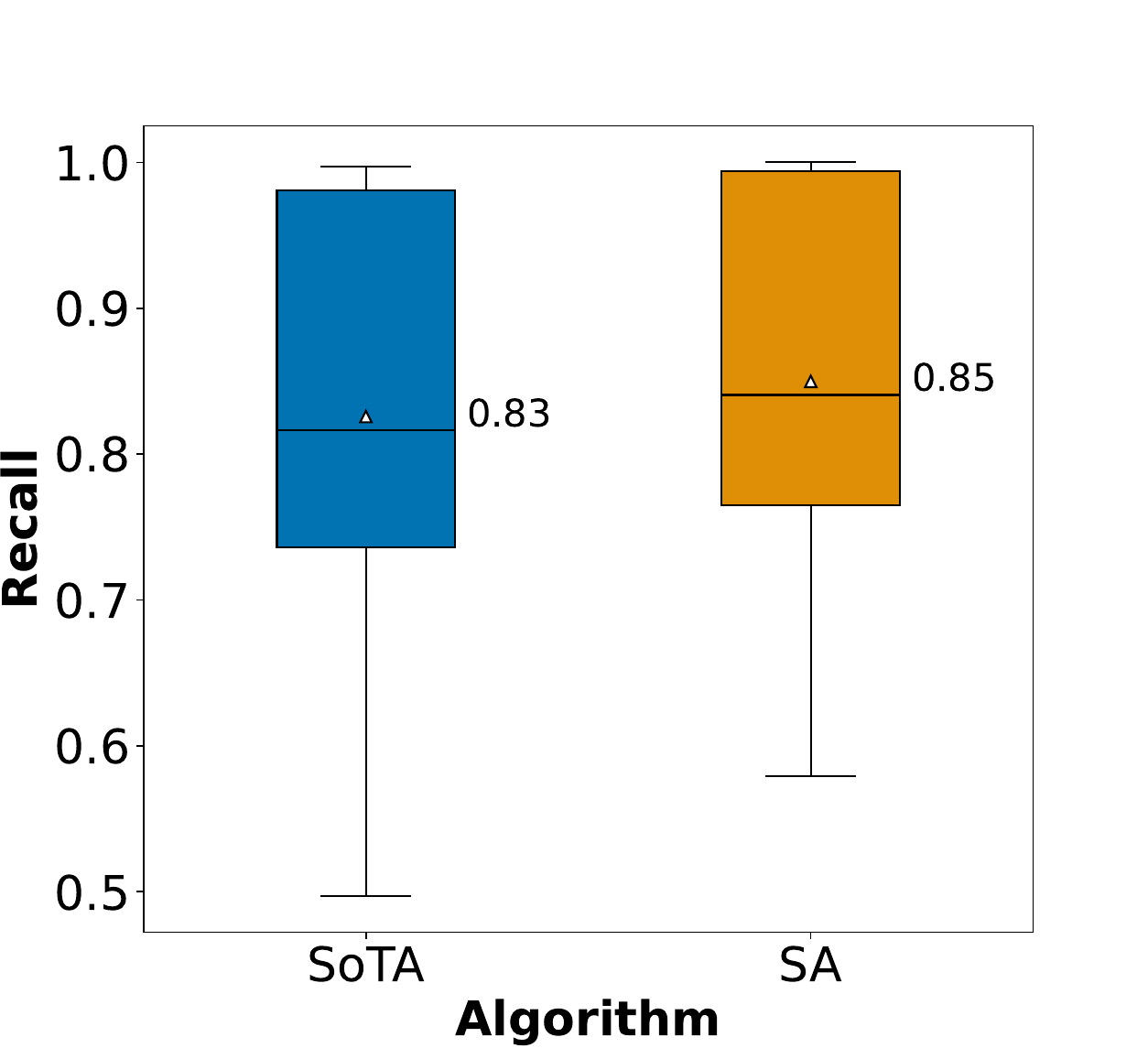} }}%
    \qquad
    \subfloat{{\includegraphics[width=6cm, height = 6cm]{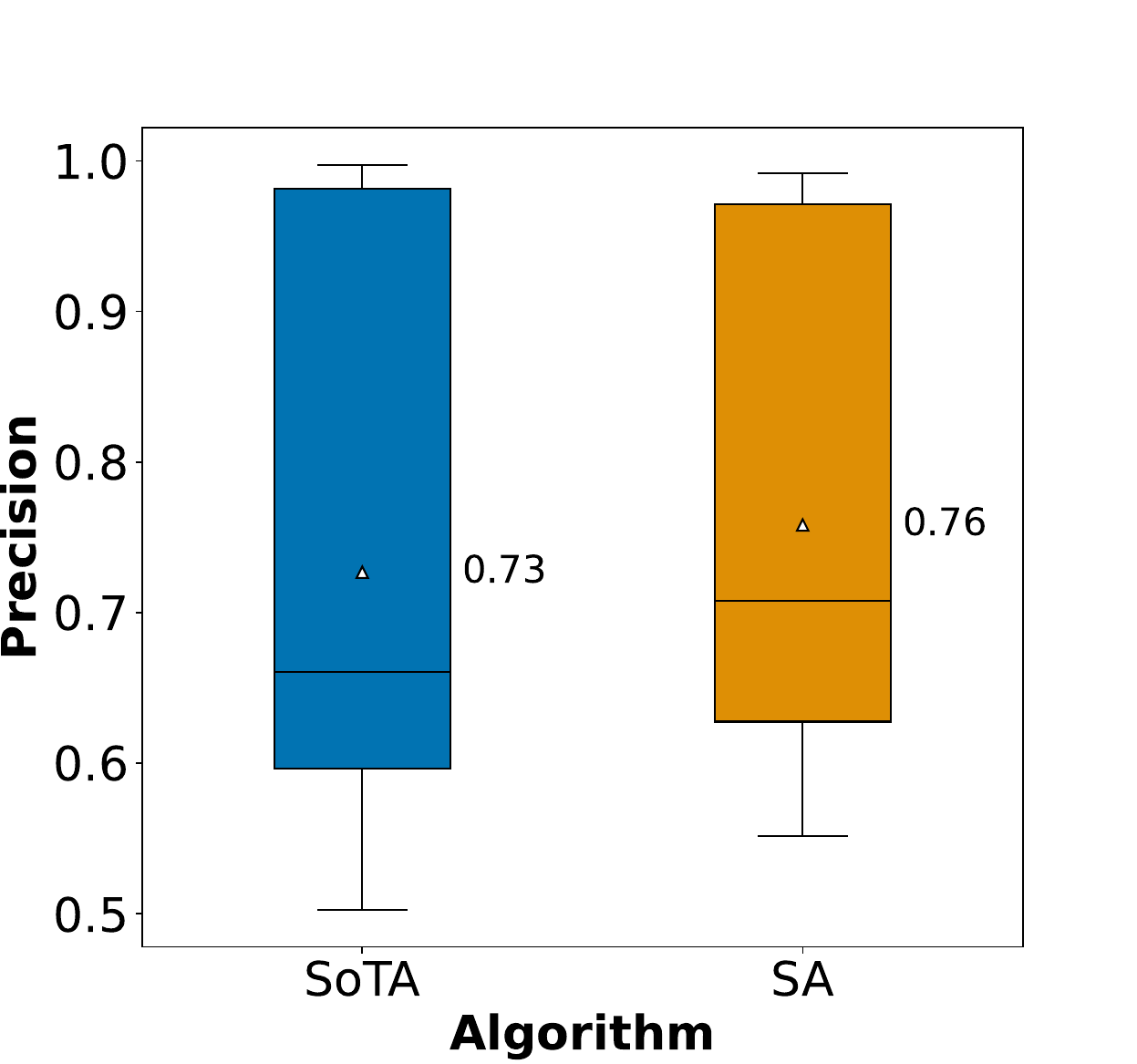} }}%
    \caption{Recall and Precision for the decision trees obtained based on \element{SA} and \element{SoTA} for four subjects in Table~\ref{tab:simulinkmodels}.}%
    \label{fig:precisionrecall-SAvsAL}%
\end{figure}

\subsection{RQ4-Usefulness} 
\label{subsec:rq4}
In view of the results of RQ2, we use the DRMs generated by the SA algorithm to evaluate their usefulness in identifying spurious failures. We focus on the DRMs for the CI subjects in Table~\ref{tab:simulinkmodels}, i.e., NTSS, AP1, AP2 and AP3. For these subjects, we can validate whether the inferred rules lead to genuinely spurious failures. 
For  NTSS, we had  access to an expert from industry, and for autopilot, we had detailed requirements and design documents. Recall from Section~\ref{subsec:fms} that the rules we obtain from DRMs are in the form of \element{IF-condition-THEN-prediction}. Each rule has a confidence that shows what percentage of the tests satisfying the condition of the rule conform to the rule's prediction label. From the DRMs for each of the four CI subjects, we extract the rules that predict the fail class with a confidence of 100\%. These rules are candidates for specifying spurious failures, since they identify conditions that lead to and only to failures. We then select the rules that are not subsumed by others through logical implication.
We use the Z3 SMT solver~\cite{z3} to find logical implications. In the end, we obtain seven rules for  NTSS, 17 rules for AP1, 24 rules for AP2 and 15 rules for AP3. On average, the rules for NTSS include two variables and three predicates, and the rules for autopilot include three variables and four predicates. 

To validate the rules for NTSS, we presented the rules to a domain expert.  Our domain expert for NTSS  is a seasoned network technologist and software engineer with more than 25 years of experience. The expert has been using the core enabling technology of NTSS (CAKE~\cite{cakepaper}, discussed earlier in this section) in commercial networking solutions since 2018. Among the seven rules for NTSS, one rule constrains an input feature formulating a sum of the NTSS input variables.   The rest of the rules involve predicates over individual input variables. The domain expert approved all the rules as they all correspond to situations where NTSS is overloaded with large traffic volumes, and hence, poor network quality is expected. We note that, for each input variable of NTSS, there is a threshold that is determined by the NTSS setup configuration. Among the seven rules, six constrain input variables near these known threshold values. An example of such rules is: \element{class6 $\geq$ $91\%\cdot$ thresh6 $\land$ class7 $\geq$ $87\%\cdot$ thresh7 THEN FAIL}  where \element{thresh6} and \element{thresh7} are thresholds of \element{class6} and \element{class7}, respectively. These rules matched the expert's intuition; nonetheless, the expert still found the rules helpful as they provide data-driven evidence for what the expert could only estimate based on experience and ad-hoc observations rather than systematically collected data. More importantly, the rule constraining a sum of input variables (as discussed in Section~\ref{sec:motivation}) was of particular interest. For this rule, neither could the domain expert estimate the combination of the variables in the rule nor was the limit in the rule close to known thresholds. This rule prompted an investigation into the source code of CAKE. This investigation confirmed that classes 7, 6 and 5 have higher priority than other classes. As such, the combined cumulative usage of these three classes needs to be capped to maintain high quality of experience. The expert indicated that, without our approach, he would not have been able to formulate such a rule merely based on his existing test scenarios and expert knowledge. 



For the autopilot's three requirements  (\element{AP1}, \element{AP2} and \element{AP3}) , we could conclusively confirm that 47 out of the 56 inferred rules represent spurious failures as per  the handbook of the De Havilland Beaver aircraft~\cite{autopilothandbook}. For the remaining nine rules, we could not ascertain whether they represent spurious failures. These rules either fail due to real faults in the system or they are spurious, but further expertise is required to confirm their spuriousness. The full set of rules and our analysis are in our supplementary material~\cite{tables20to23}.

Our results indicate that none of the 47 rules confirmed as inducing spurious failures for autopilot  could be obtained solely from the datasets generated by the preprocessing phase of SA (see Figure~\ref{fig:fig1}). Similarly, only two of the seven rules for NTSS could be obtained using the datasets from the preprocessing phase. Furthermore, from the preprocessing phase datasets alone, no additional candidate rules could be inferred for spurious failures
in either NTSS or autopilot. These findings highlight the importance of utilizing 
surrogate-assisted test generation in order to obtain useful and more comprehensive
 rules for spurious failures.

\begin{framed}
\noindent Our validation of failure-inducing rules against domain knowledge indicates that our dynamic surrogate-assisted approach is effective for identifying spurious failures.  Indeed, our results show that expert judgement alone or tests generated without the assistance of surrogates would miss many rules that one would be able to identify using our proposed approach. 
\end{framed}

\subsection{Threats to Validity} 
\label{subsec:threats}
The most important threats concerning the validity of our experiments are related to the internal and external validity.

\subsubsection{Internal Validity}

The \textit{internal validity} risks are related to confounding factors.  The effectiveness of failure-inducing rules  inferred by our approach depends on the accuracy of fitness functions and the quality of the input datasets. For Simulink models, we use an automated and provably sound technique to obtain fitness functions from logical specifications~\cite{menghi2019generating}. However, the translation of natural-language requirements into logical specifications remains a manual task and necessitates domain-expert validation. In our experiments, we mitigated the risks associated with the accuracy of fitness functions as follows: For Simulink models, the fitness functions are automatically obtained from logical specifications approved by the engineers who developed the benchmark Simulink models~\cite{nejati2019evaluating}.  For the NTSS case study, we validated the fitness function with our domain expert~\cite{enrich}. As for the risks related to the quality of datasets, we note that the labels for the data points are computed  based on the actual system outputs, and hence, are always accurate. Further,  we used adaptive random testing in the preprocessing step (see Figure~\ref{fig:fig1}) to diversify the generation of the datasets in the search input space.\\
To address any concerns regarding our comparison with \element{SoTA} in RQ3, we conducted a comprehensive review of the \element{SoTA} code~\cite{alhazengithub} to  ensure that the workflow of Algorithm~\ref{alg:alhazen}, our adaptation of \element{SoTA}, conforms to the original \element{SoTA} code. In addition, we employed the same hyper-parameters for decision trees in Algorithm~\ref{alg:alhazen} as those used by \element{SoTA}. 
Finally,  we have disclosed the code of  Algorithm~\ref{alg:alhazen} in our GitHub repository~\cite{alhazenimplementationNTSS, alhazenimplementationSimulink} to facilitate further replication and comparison efforts.

\subsubsection{External Validity} 

The subjects we used for our  evaluation and the characteristics of these subjects may influence the generalizability of our results. Related to
this threat, we note that: First, the Simuilnk models in  Lockheed's benchmark represent realistic and representative CPS components from different domains. This benchmark has been previously used in the literature on testing CPS models~\cite{nejati2019evaluating,GIANNAKOPOULOU2021106590}. 
Second, our case studies are drawn from two different domains: CPS and networks. Third, our network case study (NTSS) represents an industrial  system for which we could interact with a domain expert. The above being said, our work would benefit from  further experiments with a broader class of systems. 


\section{Lessons Learned}
\label{sec:discuss}
\emph{Lesson 1. Decision rules are a better choice than decision trees for building failure models.} In this paper, we focused on interpretable ML techniques for building failure models.
Using these techniques, we were able to generate constraints on system inputs that are easily understandable to humans~\cite{interpretablemodels}. Among   interpretable ML techniques, decision rules and decision trees have been previously used in the literature for inferring rules pertaining to a specific behaviour of a system~\cite{brindescu2020planning, ghotra2015revisiting, haq2021can, miningassumption, kampmann2020does}. We chose to use decision-rule models in our work for the following reasons: Decision rules are known to produce fewer and more concise rules compared to decision trees, which often generate many rules involving several variables and predicates. Further, decision trees are prone to the replicated subtree problem~\cite{dataminingbook}. This problem arises when the same subtree, with identical predicates and splits, appears multiple times in the tree. Replicated subtrees can increase model complexity, lead to overfitting, and hinder interpretability. Decision rules do not typically suffer from this problem, thus generally yielding more interpretable and less redundant rules. As mentioned in Section~\ref{subsec:rq4}, the rules we obtain for NTSS and autopilot, on average, have three and four predicates over two and three variables, respectively. While one could argue that limiting the depth of a decision tree, as done by the \element{SoTA} baseline, would result in a reasonably small tree, our findings indicate that, decision trees built using the parameters of the \element{SoTA} baseline lead to a $40\%$ higher number of rules and $10\%$ more predicates compared to decision rules.








\emph{Lesson 2. To evaluate a test generation algorithm for systems with numeric inputs, the accuracy and usefulness of the failure models produced by the algorithm offer more realistic insights about the algorithm than the number of individual failures found by the algorithm.} 
For systems with numeric inputs, slight modifications to the inputs of a failure-revealing test may lead to redundant failures, i.e., failures caused by the same fault. Even when one considers input diversity, e.g., measured by the Euclidean distance between test-input vectors, one cannot determine 
whether failures are non-redundant or valid by merely analyzing individual test inputs. Consequently, evaluating testing algorithms solely based on their ability to generate failures may result in misleading conclusions. Indeed, had we premised our evaluation on the number of detected failures, we would have inferred that our dynamic surrogate-assisted algorithm produces $2.3$ times more failures compared to the ML-guided and the baseline algorithms. While this conclusion would strongly favour our approach, we do not believe that this large margin is an accurate representation of the degree of improvement that our algorithm delivers. Based on the results of RQ2 and RQ3, the dynamic surrogate-assisted algorithm, when compared to alternatives, leads to an accuracy improvement \hbox{ranging from $2$\% to $14$\%.} 

\section{Related Work}
\label{sec:related}


Below, we discuss the related work on (a) the applications of ML in automated testing, (b) generating failure-inducing rules 
 and (c) test input validation. 

\textbf{Applications of Machine learning (ML) in automated testing.} ML has been widely used to enhance the effectiveness of fuzz testing~\cite{miller1990empirical} and search-based testing (SBT)~\cite{harman2009theoretical}.  In fuzz testing, ML has been employed to improve, among other things, seed generation, test sampling, and mutation-operator selection~\cite{harman2010optimizing,wang2020systematic}. In SBT, surrogate models developed based on ML have been used to effectively  and efficiently test CI systems such as cyber-physical-system controllers and simulators~\cite{matinnejad2014mil, arrieta2017search, humeniuk2021data,menghi2020approximation}, and autonomous-driving systems~\cite{ben2016testing, beglerovic2017testing, humeniuk2022search}. These approaches demonstrate that using ML can improve  the ability and the efficiency of testing in revealing faults. The ultimate goal of these approaches is to generate specific test cases. As such, these approaches are evaluated based on the number of failure-revealing tests and the severity of the failures, as determined by the fitness-function values.

Recent studies suggest that the focus of SBT should shift from generating a limited number of specific test cases to learning models that can explain system failures~\cite{feldt2020flexible}. These models can then be employed for generating multiple test cases with specific properties. Motivated by these observations, our goal is to learn failure models and focus on improving the accuracy of these models for identifying spurious failures, rather than maximizing the number and severity of failure-revealing tests, which may not accurately reflect the context where many tests fail due to spurious reasons.

\textbf{Generating failure-inducing rules.} Grammar-based test generation~\cite{hanford1970automatic} has been shown to be effective for avoiding spurious failures in fuzz testing.  More recently, grammars and probabilistic variations of grammars have been used to infer abstract failure-inducing rules~\cite{kapugama2022human}.  These rules can assist with the diagnosis of system failures, serve as accurate and high-level test oracles, and enable programmers to validate their fixes and prevent overfitting~\cite{kapugama2022human, gopinath2020abstracting, kampmann2020does}.  Our work takes inspiration from the research on inferring failure-inducing rules, but differs from the existing work on this topic in important ways. First, we focus on systems with numeric inputs, whereas existing research primarily deals with string-based inputs governed by a  grammar. Second, instead of relying on an input grammar to generate tests, we investigate various test-generation heuristics that are guided by quantitative fitness functions drawn from system requirements. An exception  is the work of B\"{o}hme et al.~\cite{bohme2020human}, which infers program patches for numeric systems without the need for input grammars. However, this approach relies on the availability of a human oracle to validate the verdicts of individual test inputs. In our context, this would be expensive and likely infeasible. Our work further differs from the above in that our goal is to identify rules for spurious failures rather than generating program patches.



 The closest work to ours is the Alhazen framework~\cite{kampmann2020does}, which we compared with in RQ3.   In addition to the discussion and empirical comparison  in RQ3, we note that our approach differs from Alhazen in its input-feature engineering for failure models. We 
 derive the input features for decision-rule models from domain-knowledge heuristics, whereas Alhazen derives the input features dynamically from its input grammar.  While Alhazen automates  input-feature engineering, by incorporating domain knowledge into the design of  input features, our approach provides the flexibility to derive rules that  more closely match expert intuition.

\textbf{Test Input Validation.}  
Test input validation determines whether the test inputs given to a system  adhere to the format, range or constraints anticipated by the system  requirements. Test input validation improves the reliability of  test results and is crucial for software testing~\cite{whenandwhy}. Traditional software testing and verification approaches for CPS and network systems assume that pre-conditions describing valid inputs are already specified~\cite{menghi2020approximation, menghi2019generating, staliro} or, alternatively, rely on formal assume-guarantee and design-by-contract techniques~\cite{derler2013cyber,sangiovanni2012taming,henzinger1998you}. Techniques based on assume-guarantee and design by contract require high-level formal system specifications. Such specifications  do not exist for many real-world systems including our study subjects. 
Recent studies explore test input validity for deep learning (DL) models~\cite{ifthehuman,whenandwhy},  demonstrating that existing DL testing techniques generate several invalid test inputs. To mitigate this problem,  the studies preform human subject experiments and establish metrics that  determine test input validity for DL models. 
Although our work is not concerned with DL models, our definition of  spurious test inputs is similar to 
that of invalid inputs for DL models~\cite{whenandwhy,ifthehuman}. Similar to DL testing, failing to account for input validity leads to the generation of many invalid test inputs, thus reducing the reliability of test generation. In addition, similar to the research for DL models, we identify the rules leading to spurious failures for our study subjects based on domain expertise and human knowledge.


\section{Conclusion}
\label{sec:conclusion}
In this paper, we presented a  data-driven framework for inferring failure models for systems with numeric inputs including cyber-physical and network systems. The framework employs existing surrogate-assisted and machine learning-guided (ML-guided) test generation techniques. We proposed a new dynamic surrogate-assisted algorithm that uses multiple surrogate models simultaneously during search, and dynamically selects the predictions from the most accurate model.
We compared the accuracy of failure models obtained using our dynamic surrogate-assisted approach against two ML-guided techniques as well as two baselines using $16$ study subjects from the cyber-physical and network domains.  Our results,  confirmed by statistical tests, show that the average accuracy, precision and recall of the dynamic surrogate-assisted approach are higher than those of the ML-guided test generation algorithms, and of the state-of-the-art and random-search baselines. Moreover, the rules inferred from the failure models built for our compute-intensive subjects identify genuine spurious failures as validated against domain knowledge. For future work, we plan to apply our approach to computer-vision and autonomous-driving systems and subsequently use the rules inferred by failure models as guidance for generating test inputs.


\section{Data availability}
\label{sec:data}
Implementations of all algorithms are available at~\cite{sourcecode}. The Simulink benchmark is available at~\cite{simulinkbenchmark}, and NTSS at~\cite{enrichgithub}. The requirements specifications for all study subjects are provided at~\cite{requirements}. Our evaluation data includes: (1)~raw datasets for the experiments~\cite{datasets}; (2)~rules generated for CI subjects~\cite{APNTSSrules, tables20to23}; (3)~evaluation scripts~\cite{evaluationcode} and the analyzed data~\cite{evaluationresults}; (4)~statistical analysis results~\cite{statisticalgithub, tables8to13, table5}; and (5)~scripts for the plots in the paper~\cite{evaluationcode}. 

\section*{Acknowledgements}
We gratefully acknowledge the financial support received from NSERC of Canada through their Alliance, Discovery, and Discovery Accelerator programs.

\bibliographystyle{ACM-Reference-Format}
\bibliography{bibliography}

\end{document}